\documentclass[preprint,authoryear,sort&compress,12pt]{elsarticle}  % final
\usepackage[top=1.25in, bottom=1.25in, left=1.25in, right=1.25in]{geometry}
\usepackage{graphicx}
\usepackage{amssymb}
\usepackage{amsmath}
\usepackage{lineno}
\usepackage{subfig}

\usepackage{color}
\usepackage{soul}
\usepackage{array}
\usepackage{hyperref}
\usepackage[colorinlistoftodos]{todonotes}

\newcommand{\xmb}[1]{\ensuremath{\mathbf{#1}}}
\newcommand{\xmbs}[1]{\ensuremath{\boldsymbol{#1}}}

\newcommand*\patchAmsMathEnvironmentForLineno[1]{%
  \expandafter\let\csname old#1\expandafter\endcsname\csname #1\endcsname
  \expandafter\let\csname oldend#1\expandafter\endcsname\csname end#1\endcsname
  \renewenvironment{#1}%
     {\linenomath\csname old#1\endcsname}%
     {\csname oldend#1\endcsname\endlinenomath}}% 
\newcommand*\patchBothAmsMathEnvironmentsForLineno[1]{%
  \patchAmsMathEnvironmentForLineno{#1}%
  \patchAmsMathEnvironmentForLineno{#1*}}%
\AtBeginDocument{%
\patchBothAmsMathEnvironmentsForLineno{equation}%
\patchBothAmsMathEnvironmentsForLineno{align}%
\patchBothAmsMathEnvironmentsForLineno{flalign}%
\patchBothAmsMathEnvironmentsForLineno{alignat}%
\patchBothAmsMathEnvironmentsForLineno{gather}%
\patchBothAmsMathEnvironmentsForLineno{multline}%
}
\journal{International Journal of Multiphase Flow}

\begin{document}

\begin{frontmatter}

  \title{Diffusion-Based Coarse Graining in Hybrid Continuum--Discrete Solvers: Applications in
    CFD--DEM}

 \author{Rui Sun} \ead{sunrui@vt.edu}
 \author{Heng Xiao\corref{corxh}} \ead{hengxiao@vt.edu}
 \address{Department of Aerospace and Ocean Engineering, Virginia Tech, Blacksburg, VA 24060, United
 States}

 \cortext[corxh]{Corresponding author. Tel: +1 540 231 0926}

\begin{abstract}
  In this work, a coarse-graining method previously proposed by the authors in a companion paper
  based on solving diffusion equations is applied to CFD--DEM simulations, where coarse graining is
  used to obtain solid volume fraction, particle phase velocity, and fluid--particle interaction
  forces.  By examining the conservation requirements, the variables to solve diffusion equations
  for in CFD--DEM simulations are identified. The algorithm is then implemented into a CFD--DEM solver
  based on OpenFOAM and LAMMPS, the former being a general-purpose, three-dimensional CFD solver
  based on unstructured meshes.  Numerical simulations are performed for a fluidized bed by using
  the CFD--DEM solver with the diffusion-based coarse-graining algorithm. Converged results are
  obtained on successively refined meshes, even for meshes with cell sizes comparable to or smaller
  than the particle diameter. This is a critical advantage of the proposed method over many existing
  coarse-graining methods, and would be particularly valuable when small cells are required in part
  of the CFD mesh to resolve certain flow features such as boundary layers in wall bounded flows and
  shear layers in jets and wakes. {\color{black} Moreover, we demonstrate that the overhead
    computational costs incurred by the proposed coarse-graining procedure are a small portion of
    the total computational costs in typical CFD--DEM simulations as long as the number of particles
    per cell is reasonably large, although admittedly the computational overhead of the
    coarse-graining procedure often exceeds that of the CFD solver.} Other advantages of the
  diffusion-based algorithm include more robust and physically realistic results, flexibility and
  easy implementation in almost any CFD solvers, and clear physical interpretation of the
  computational parameter needed in the algorithm. In summary, the diffusion-based method is a
  theoretically elegant and practically viable option for practical CFD--DEM simulations.
\end{abstract}


\end{frontmatter}

% \linenumbers

\section{Introduction}
\label{sec:intro}

\subsection{Particle-Laden Flows: Physical Background and Multi-Scale Modeling}
\label{sec:flow}
Particle-laden flows occur in many settings in natural science and engineering, e.g., sediment
transport in rivers and coastal oceans, debris flows during flooding, cuttings transport in
petroleum-well drilling, as well as powder handling and pneumatic conveying in pharmaceutical
industries~\citep{sifferman74dc, nielsen92cb, iverson97ph, yang98fs}. In particular, this work
is concerned with the dense-phase regime, where both the fluid--particle interactions and the
inter-particle collisions play important roles. For examples, the dense particle-laden flows in
fluidized beds in chemical reactors and the sheet flows in coastal sediment transport both fall
within this regime.

Various numerical simulation approaches have been proposed for particle-laden flows in the past few
decades. Among the most established and most commonly used is the Two-Fluid Model (TFM) approach,
which describes both the fluid phase and the particle phase as inter-penetrating
continua~\citep{sun2007hybrid}. The two sets of mass and momentum conservation equations for the two
phases are solved with mesh-based numerical discretization, with coupling terms accounting for the
interaction forces between the phases. Particles are not explicitly resolved or represented in the
TFM formulation, although the particle phase properties do take into account certain particle
characteristics. Therefore, the computational cost of the two-fluid model is relatively low, and
thus this method is widely used in industrial applications, where fast turnover times are often a
critical requirement. However, the physics of the particle or granular flows are fundamentally
different from that of fluids. Among many other difficulties associated with the TFM, a
critical issue with this approach is that a universal constitutive relation for the particle phase
that is applicable to different flow regimes seems to be lacking despite much research on this
topic~\citep{sun11jfm}.  This difficulty stems from the fact that unlike the flow of real continuum
fluids (gases or liquids) where strong separation of scales justifies the continuum description, in
granular flows the scale separation is weak~\citep{glasser01scale}, i.e., the representative volume
element can be of a similar order of magnitude to particle diameters. As such, a continuum
description of the particle phase would suffer from these intrinsic difficulties. Other drawbacks of
the TFM approach include the difficulty in representing particles with a continuous distribution of
diameters or densities and the reliance on empirical models of fluid--particle interactions, among
others~\citep{sun09}.

On the other hand, direct numerical simulations based on Lattice-Boltzmann method~\citep{yin08} or
by solving Navier--Stokes equations with fluid--particle interfaces fully resolved (e.g., via
immersed boundary method~\citep{Kempe:14on}) are computationally expensive. The DNS methods are
currently limited to systems of \( O(10^3) \) particles in spite of sustained rapid growth of
available computational resources in the past decades. Interestingly, this difficulty is also due to
the fact that multiple scales do exist in the particle-laden flow problem, although the scale
separation is weak, as explained above. That is, the scales of concern are several orders of
magnitude larger than the particle diameter $d_p$, and thus the simulated system may contain a large
number of particles. It is expected that DNS will not be affordable for simulating realistic dense
particle-laden flows in the near future, where the number of particles can be \( O(10^6) \) or even
more.

In view of the multi-scale nature of and the weak scale-separation in dense particle-laden flows,
the continuum--discrete approach seems to be a natural choice. In this approach, continuum model is
used to describe the fluid phase, while the particle phase is described by the Discrete Element
Method (DEM) , where particles are tracked individually based on Newton's second law in a Lagrangian
framework. DEM was first used to model granular flow without interstitial fluids in geotechnical
engineering in the 1970s~\citep{cundall79}. The hybrid CFD--DEM approach to model particle-laden
flows was attempted in the 1990s~\citep{tsuji93}. Traditionally the locally averaged Navier--Stokes
equations are adopted as the continuum model~\citep{anderson67}, leading to a hybrid method commonly
referred to as CFD--DEM (Computational Fluid Dynamics--Discrete Element Method). Recently, Large
Eddy Simulation (LES), a CFD technology based on the solution of filtered Navier--Stokes equations,
has been used as the continuum fluid model, leading to hybrid LES--DEM
solvers~\citep{Zhou2004}. Other variations in the category of continuum--discrete solvers include
those using Smooth Particle Hydrodynamics (SPH) or Lattice-Boltzmann for the fluid
flow~\citep{han07ns, sun2013three}.

\subsection{Coarse Graining in Continuum--Discrete Particle-Laden Flow Solvers}
\label{sec:need-cg}

In all these continuum--discrete particle-laden flow solvers mentioned above including CFD--DEM and
LES--DEM, one needs to bridge the continuum-based conservation equations for the fluid phase and the
discrete description of the particle phase. Specifically, the presence and the dynamic effects of
the particles on the fluid are taken into account in the fluid continuity and momentum equations
through the macroscopic quantities of the particle phase, e.g., solid volume fraction \(
\varepsilon_s \), solid phase velocity \( \xmb{U}_s \), and solid--fluid interphase forces
$\xmb{F}^{fp}$. These Eulerian field quantities are \emph{not solved for} in the continuum-scale
solver, but need to be obtained from the discrete particle information (i.e., individual particle
locations \( \xmb{x}\), particle velocities \( \xmb{u} \), interaction forces on individual
particles \(\xmb{f}^{fp} \)). The process of obtaining macroscopic quantities from particle-scale
quantities is referred to as \emph{coarse graining} in this work.

In CFD--DEM or LES--DEM solvers the fluid equations are discretized with mesh-based numerical
methods such as finite volume for finite element methods. From here on we focus our discussions on
CFD--DEM for brevity. However, note that the discussions presented and the methods proposed in this
work shall be equally applicable to LES--DEM solvers, and may be useful for other
continuum--discrete methods such as SPH--DEM and LB-DEM for particle-laden flows. Another method
that is closely related to CFD/LES--DEM is the Particle-in-Cell (PIC) method, which is widely used
in plasma simulations~\citep{dawson1983particle}, where individual physical particles (electrons,
ions, etc.)  or ``super-particles'' that represent a number of physical particles of similar
properties are tracked in a Lagrangian framework. The interactions among the particles are computed
not in a pair-wise way but via electric and magnetic fields that are Eulerian field quantities
computed from the particle distribution data. The coarse graining is an important ingredient in the
PIC method, and the proposed method can be of relevance there.

Details on how the solid phase quantities interact with the fluid phase quantities in CFD--DEM will
be presented in Section~\ref{sec:cfddem} after the mathematical formulation for the method is
introduced. In CFD--DEM solvers the Eulerian field quantities of the fluid phase become cell-based
quantities after numerical discretization. Therefore, to bridge continuum-based description of the
fluid phase and the discrete description of the particle phase, we simply need to obtain cell-based
representation of the Eulerian field quantities (e.g., solid volume fraction \( \varepsilon_s \),
Eulerian velocity \( \xmb{U}_s \), and fluid--particle interaction forces \( \xmb{F}^{fp} \)) of the
solid phase. A straightforward and probably the most widely used method to link particle quantities
and cell quantities is the Particle Centroid Method (PCM). The PCM utilizes the fluid mesh for
coarse graining by summing over all particle volumes in each cell to obtain cell-based solid volume
fraction \( \varepsilon_s\), and similar procedures are followed for other variables such as \(
\xmb{U}_s \) and \( \xmb{F}^{fp} \). This method is very straightforward to implement in almost any
CFD solvers, but it can lead to large errors when cell size to particle diameter ratios are
small. Consequently, various alternatives have been proposed to improve the accuracy of PCM.
{\color{black} The Divided Particle Volume Method (DPVM), first proposed and implemented
  by~\cite{wu09three,wu09accu}, is such an example. In this method, the volume of a particle is
  divided among all cells that it overlaps with according to the portion of the volume within each
  cell, and is not only distributed entirely to the cell its centroid resides in as in PCM. As a
  consequence, the DPVM at least guarantees that the solid volume fraction $\varepsilon_s$ in any
  cell should never exceed one, effectively preventing very large gradients in the obtained
  $\varepsilon_s$ field. DPVM works for arbitrary meshes, structured or unstructured, with any
  elements shapes as long as any edge of the cell has a length larger than the particle
  diameter. Comprehensive comparisons between DPVM and PCM recently performed by~\citet{peng14in}
  suggest that DPVM has significantly improved performance over PCM.} Another idea, recently
proposed by~\cite{Deb13-ano}, is to use two separate meshes for the CFD discretization and the
coarse graining.  While the improved variants do outperform the PCM in terms of accuracy, the
implementations of these sophisticated methods are often significantly more complicated, especially
in CFD solver based on unstructured, non-Cartesian meshes.

In our efforts to develop a CFD/LES--DEM solver with a parallel, three dimensional CFD code based on
unstructured meshes with arbitrary cells shapes, we found that none of existing coarse-graining
methods is able to satisfy the requirements of easy implementation and good accuracy
simultaneously. The difficulties motivated us to develop a coarse-graining method that is suitable
for practical implementation in general-purpose CFD/LES--DEM solvers, while maintaining the
theoretical rigor and excellent accuracy. 

{\color{black} The general motivation, description, and derivation of the diffusion-based
  coarse-graining method as well as \emph{a priori} tests (where no CFD--DEM simulations were performed)
  have been presented in~\cite{part1}.  Specifically, the companion paper (1) comprehensively
  reviewed and compared existing coarse-graining methods in the literature, including PCM, DPVM,
  two-grid formulation, and statistical kernel methods, (2) motivated and proposed a diffusion-based
  coarse-graining method, (3) demonstrated the equivalence (up to the mesh discretization accuracy)
  between the current method and the statistical kernel-based coarse-graining method with Gaussian
  kernel, and (4) evaluated the performance of the diffusion-based method by comparing it with 
  existing methods in various scenarios, with both structured and unstructured meshes, and both in
  the interior domain and near wall boundaries.  While maintaining all the merits of its
  theoretically equivalent counterpart such as mesh-independence, the diffusion-based method is much
  easier for practical implementations in general-purpose CFD--DEM solvers, and provides a unified
  framework for treating interior particles and particles that are located near boundaries.

  The present work is a companion paper of~\cite{part1}. The objective is to explore the
  theoretical and practical issues of applying the diffusion-based coarse-graining method in a
  general-purpose CFD--DEM solver, and to evaluate its performance in practical fluidized bed
  simulations.  Specifically, in this paper (1) the conservation characteristics of the
  diffusion-based coarse-graining method are studied, based on which the variables to solve
  diffusion equations for are identified (i.e., $\varepsilon_s$, $\varepsilon_s \mathbf{U}_s$,
  $\varepsilon_f \mathbf{F}^{fp}$), (2) the algorithm is implemented into a CFD--DEM solver and
  tested in fluidized bed simulations, highlighting the improved mesh-convergence behavior compared
  to the PCM, and (3) the choice of diffusion bandwidth is justified based on physical reasoning.
  The issues discussed in the present work (e.g, the in-situ performance of the proposed
  coarse-graining method in CFD--DEM solvers, as well as the choice of variables to solve diffusion
  equations for and the diffusion bandwidth) are specific to the application of the diffusion-based
  method in CFD--DEM simulations. These issues are not trivial and warrant thorough
  investigations. }

The rest of the paper is organized as follows. Section~\ref{sec:cfddem} introduces the mathematical
formulation of the CFD--DEM approach, gives a summary of the diffusion-based coarse-graining method,
and then discusses their numerical implementations and the numerical methods used in the
simulations.  In Section~\ref{sec:insitu} CFD--DEM simulations are conducted by using the proposed
coarse-graining method, and the results are discussed and compared with those obtained with PCM.
The overhead computational costs associated with the coarse-graining procedure are investigated in a
series of cases with different ratios of particle and cell numbers.  The physical basis of choosing the
bandwidth parameter in the diffusion-based method and possible extensions to spatial--temporal
averaging are discussed in Section~\ref{sec:discuss}. Finally, Section~\ref{sec:conclude} concludes
the paper.

\section{Methodology}
\label{sec:cfddem}

\subsection{Mathematical Formulations of CFD--DEM}

Due to the large number of symbols and subscripts used in this paper, it is beneficial to establish
certain conventions in the notations before proceeding to the presentation of the particle and fluid phase
equations. Unless noted otherwise, superscripts are used to categorize the physical background
associated with a quantity, e.g., `$col$' for collision, `$fp$' for fluid--particle interactions, etc.
These superscripts should be relatively self-evident. Phase subscripts are used to denote
quantities associated with solid phase (`$s$'),  fluid phase (`$f$'), individual particles (`$p$'),
and individual cells (`$c$'). Index subscripts $i$ and $k$ are used as indices for particles and
cells, respectively. To avoid further cluttering of indices, vector notations are preferred to tensor
notations throughout the paper. When a quantity has both the indices ($i$ or $k$) and phase
subscripts ($s$, $f$, $p$, or $c$), they are separated by a comma. The particle-level velocities
and the forces associated with individual particles are denoted as \(\xmb{u}\) and \( \xmb{f} \),
respectively; and the velocities and forces in the continuum scale are denoted as \( \xmb{U} \) and
\(\xmb{F} \), respectively.

\subsubsection{Discrete Element Method for Particles}
\label{sec:dem-particle}

In the CFD--DEM approach, the translational and rotational motion of each particle is described by
the following equations~\citep{cundall79,ball97si,weber04}: 
\begin{subequations}
 \label{eq:newton}
 \begin{align}
  m \frac{d\xmb{u}}{dt} &
  = \xmb{f}^{col} + \xmb{f}^{fp} + m \xmb{g} \label{eq:newton-v} \\
  I \frac{d\xmbs{\Psi}}{dt} &
  = \xmb{T}^{col} + \xmb{T}^{fp} \label{eq:newton-w}
 \end{align}
\end{subequations}
where \( \xmb{u} \) is the particle velocity; $t$ is time; $m$ is particle mass; \(\xmb{f}^{col} \)
is the force due to collisions and enduring contacts with other particles or wall boundaries;
\(\xmb{f}^{fp}\) denotes the forces due to fluid--particle interactions, e.g., drag, lift, and
buoyancy; \(\xmb{g}\) denotes external body forces. Similarly, \(I\) and \(\xmb{\Psi}\) are angular
moment of inertia and angular velocity, respectively, of the particle; \(\xmb{T}^{col}\) and
\(\xmb{T}^{fp}\) are the torques due to particle--particle interactions and fluid--particle
interactions, respectively. For the purpose of computing collision forces and torques, the particles
are modeled as soft spheres with interparticle contact represented by an elastic spring and a
viscous dashpot.  Further details can be found in the
literature~\citep[e.g.,][]{cundall79,tsuji93,xiao-cicp}.

\subsubsection{Locally-Averaged Navier--Stokes Equations for Fluids}
\label{sec:lans}

The fluid phase is described by the locally averaged incompressible Navier--Stokes equations.
Assuming constant fluid density \(\rho_f\), the continuity and momentum equations for the fluid are~\citep{anderson67,kafui02}:
\begin{subequations}
 \label{eq:NS}
 \begin{align}
  \nabla \cdot \left(\varepsilon_s \xmb{U}_s + {\varepsilon_f \xmb{U}_f}\right) &
  = 0 , \label{eq:NS-cont} \\
  \frac{\partial \left(\varepsilon_f \xmb{U}_f \right)}{\partial t} + \nabla \cdot \left(\varepsilon_f \xmb{U}_f \xmb{U}_f\right) &
  = \frac{1}{\rho_f} \left( - \nabla p + \nabla \cdot {\xmbs{\mathcal{R}}} + \varepsilon_f \rho_f \xmb{g} + \xmb{F}^{fp}\right), \label{eq:NS-mom}
 \end{align}
\end{subequations}
where \(\varepsilon_s\) is the solid volume fraction; \( \varepsilon_f = 1 - \varepsilon_s \) is the
fluid volume fraction; \( \xmb{U}_f \) is the fluid velocity. The four terms on the right hand side
of the momentum equation are pressure (\(p\)) gradient, divergence of stress tensor \(
\xmbs{\mathcal{R}} \) (including viscous and Reynolds stresses), gravity, and fluid--particle
interactions forces, respectively. Since the equations are formulated in the Eulerian framework, all
variables herein are continuum quantities, i.e., they are mesh-based when discretized
numerically. As explained in Section~\ref{sec:flow}, the solid phase quantities \( \varepsilon_s \),
\( \xmb{U}_s \), \( \xmb{F}^{fp} \) are not explicitly solved for, but are instead obtained from the
particle information via coarse-graining procedures.

\subsubsection{Fluid--Particle Interactions}
\label{sec:fpi}
While the fluid--particle interaction force \( \xmb{F}^{fp} \) consists of many components including
buoyancy \( \xmb{F}^{buoy} \), drag \( \xmb{F}^{drag} \), force, and Basset history
force, among others, here we focus on the drag term for the purpose of illustrating the bridging between the
continuum and discrete scales. Other forces can be coarse grained in a similar way. In PCM-based
coarse graining, the particle drag on the fluid is obtained by summing the drag over all particle in
a cell. The drag on an individual particle $i$ is generally formulated as:
\begin{equation}
  \mathbf{f}^{drag}_i = \frac{V_{p,i}}{\varepsilon_{f, i} \varepsilon_{s, i}} \beta_i \left( \mathbf{u}_{p,i} -
  \mathbf{U}_{f, i} \right),
  \label{eqn:particleDrag}
\end{equation}
where \( V_{p, i} \), and \( \mathbf{u}_{p, i} \) are the volume, and
the velocity, respectively, of particle $i$; \( \mathbf{U}_{f, i} \) is the fluid velocity
interpolated to the center of particle $i$; \( \beta_{i} \) is the drag correlation coefficient.
Various correlations have been proposed for $\beta$ in dense particle-laden flows~\citep{mfix93,di94vo,wen13me}, which account for the presence of other particles when calculating the drag on a
particle by incorporating $\varepsilon_s$ in the correlation forms. The correlation of~\cite{mfix93}
is adopted in this work. However, the specific form of the correlation is not
essential for the present discussion, and is thus omitted here for brevity. It suffices to point
out that regardless of the specific form of the drag correlations, the solid volume fraction \(
\varepsilon_{s, i} \) and the fluid velocity \( \mathbf{u}_{f, i} \) local to the particle are
needed to calculate $\beta$. Both \( \varepsilon_{s, i} \) and \( \mathbf{u}_{f, i} \) are Eulerian
mesh-based quantities interpolated to the centroid location \( \xmb{x}_i \) of particle $i$.

To summarize, in CFD--DEM simulations the following Eulerian mesh-based quantities need to be
obtained by coarse graining the particle data:
\begin{enumerate}
\item
solid volume fraction \( \varepsilon_s \),
\item
 solid phase velocity \( \xmb{u}_s \), and
\item
 fluid-particle interaction force \( \xmb{F}^{fp} \).
\end{enumerate}
 These fields are needed in solving the continuity and momentum equations (\ref{eq:NS-cont})
 and~(\ref{eq:NS-mom}) for the fluid phase. Eulerian field quantities that need to be interpolated
 to particle locations include \( \varepsilon_s \),  \( \xmb{u}_s \), and  \( \xmb{u}_f \), which
 are needed for the calculation of fluid forces on individual particles. It can been seen that the
 coarse graining and interpolation are of critical importance for modeling the interactions between
 the continuum and discrete phases.

\subsection{Diffusion-Based Coarse-Graining Method}

\subsubsection{Summary of the Diffusion-Based Method}
\label{sec:dbcg-alg}

The proposed algorithm is built upon the particle centroid method, which is the coarse-graining
method used in most CFD--DEM solvers. Therefore, here we first introduce the PCM algorithm in
detail. To calculate solid volume fraction field $\varepsilon_s$ with PCM, we loop through all cells
to sum up all the particles volume to their host cells (defined as the cell within which the
particle centroid is located), thus obtaining the total particle volume in each cell. The solid
volume fraction for cell \(k\) is then obtained by dividing the total particle volume in the cell by
the total volume of the cell \(V_{c, k}\).  That is,
\begin{equation}
 \varepsilon_{s,k}
 = \frac{\sum_{i=1}^{n_{p, k}} V_{p, i}}{V_{c, k}} \, , \label{eq:pcm-k} 
\end{equation}
where \(n_{p, k}\) is the number of particles in cell \(k\), which implies that \(\sum_{k=1}^{N_c}
n_{p, k} = N_p\).  The $\varepsilon_s$ field obtained with the PCM procedure above (denoted as
$\varepsilon_0$ for reasons that will soon be evident) may have unphysically large values for some
cells and consequently very large spatial gradients, which can cause instabilities in CFD--DEM
simulations or lead to numerical artifacts. To address this issue, in the diffusion based method we
proposed in the companion paper~\citep{part1}, a transient diffusion equation for
$\varepsilon_s(\mathbf{x}, \tau)$ is solved with initial condition $\varepsilon_0$ and no-flux
boundary conditions:
\begin{subequations}
\begin{align}
 \frac{\partial \varepsilon_s}{\partial \tau} &
 =  \nabla^2 \varepsilon_s \quad \textrm{ for } \; \mathbf{x} \in \Omega, \tau > 0
 \label{eq:diffusion}
 \\
 \varepsilon_s(\mathbf{x}, \tau=0) &
 = \varepsilon_0 (\mathbf{x}) \label{eq:diffusion-b} \\
\frac{\partial \varepsilon_s}{\partial \mathbf{n}} &
= 0  \textrm{ on } \partial \Omega
\label{eq:diffusion-c}
\end{align}
\end{subequations}
where \(\mathbf{x} \equiv [x, y, z]^T\) are spatial coordinates; $\Omega$ is the computational
domain; \(\nabla^2 \varepsilon_s= {\partial^2 \varepsilon_s}/{\partial x^2} + {\partial^2
  \varepsilon_s}/{\partial y^2} + {\partial^2 \varepsilon_s}/{\partial z^2}\) in Cartesian
coordinates; $\varepsilon_0 (\mathbf{x}) $ is the solid volume fraction field obtained with the PCM;
\(\tau\) is pseudo-time, which should be distinguished from the physical time $t$ in the CFD--DEM
formulation. Finally, $\partial \Omega$ is the boundary of $\Omega$; $\mathbf{n}$ is the surface
normal of $\partial \Omega$.  The diffusion equation~(\ref{eq:diffusion}) is integrated until time
$\tau = T$ with the initial condition Eq.~(\ref{eq:diffusion-b}) and boundary condition
Eq.~(\ref{eq:diffusion-c}), and the obtained field $\varepsilon_s(\mathbf{x}, T)$ is the solid
volume fraction field to be used in the CFD--DEM formulation.  The end time $T$ is a physical
parameter characterizing the length scale of the coarse graining.  It was demonstrated that the
diffusion-based method above is equivalent to the statistical kernel function-based coarse graining
with the following Gaussian kernel:
\begin{equation}
  h_i = h(\mathbf{x}-\mathbf{x}_i) = \frac{1}{(b^2 {\pi})^{3/2}}
 \exp \left[ - \frac{(\mathbf{x}-\mathbf{x}_i)^T (\mathbf{x}-\mathbf{x}_i) }{b^2} \right]
 \label{eq:hi}
\end{equation}
where $h_i$ is the kernel associated with particle $i$, which is located at $\mathbf{x}_i$. The
equivalence between the two is established with $b=\sqrt{4T}$. Moreover, for particles located near
boundaries (e.g., walls), the kernel-based methods need to be modified to satisfy conservation
requirements with methods such as method of images~\citep{zhu02ave}. It was further demonstrated
that the diffusion-based coarse-graining method above satisfies conservation requirements
automatically, and thus interior and near-boundary particles are treated in a unified framework.  In
fact, with the no-flux boundary conditions the diffusion-based method is equivalent to the method of
images proposed by~\citet{zhu02ave}.

The solid phase velocity \(\mathbf{U}_{s, k}\) and the fluid--particle interaction force
$\mathbf{F}_{k}^{fp}$ per unit mass in cell \(k\) are computed in a similar way. That is, the
initial fields are first obtained by using PCM:
\begin{align}
 \mathbf{U}_{s, k} &
 = \frac{\sum_{i=1}^{n_{p, k}} \rho_s V_{p, i} \, \mathbf{u}_{p, i}} {\sum_{i=1}^{n_{p, k}} \rho_s V_{p, i}}
= \frac{\sum_{i=1}^{n_{p, k}} \rho_s V_{p, i} \, \mathbf{u}_{p, i}} {\rho_s \varepsilon_{s,k} V_{c, k}} \, , \label{eq:pcm-u} \\
 \mathbf{F}_{k}^{fp} &
 = \frac{- \sum_{i=1}^{n_{p, k}} \, \mathbf{f}^{fp}_i} {\rho_f  \varepsilon_{f, k}  V_{c, k}} \, , \label{eq:pcm-f}
\end{align}
and then, diffusion equations are solved for the fields $\varepsilon_s \mathbf{U}_s $ and
$\varepsilon_f \mathbf{F}^{fp}$. After the coarse graining is performed on $\varepsilon_s
\mathbf{U}_s $ and $\varepsilon_f \mathbf{F}^{fp}$, the coarse-grained fields are divided by
$\varepsilon_s$ and $\varepsilon_f$ (i.e., $1-\varepsilon_s$), respectively, to obtain $\mathbf{U}_s
$ and $\mathbf{F}^{fp}$.  Note that solving diffusion equations directly for $\mathbf{U}_{s, k} $
and $\mathbf{F}_{k}^{fp}$ would violate conservation requirements. While $\varepsilon_{s,k}$ seems
to be the intuitive and natural choice to solve diffusion equations for, the choices of
$\varepsilon_s \mathbf{U}_s $ and $\varepsilon_f \mathbf{F}^{fp}$ as the variables to solve
diffusions for are not straightforward.  Justifications are thus provided below, and detailed proofs
are presented in the Appendix.

\subsubsection{Conservation Characteristics and Choice of Diffusion Variables}
\label{sec:conserve}
It is critical that any coarse-graining algorithm should conserve the relevant physical quantity in
the coarse-graining procedure. Specifically in the context of CFD--DEM simulations these quantities
include total particle mass, particle phase momentum, and total momentum of the fluid--particle
system. The conservation requirement implies that the total mass computed from the coarse-grained
continuum field should be the same as the total particle mass in the discrete phase. Similarly, when
calculating solid phase velocity \( \mathbf{U}_s \), the total momentum of the particles should be
conserved before and after the coarse graining; finally, to conserve momentum in the fluid--particle
system, the total particle forces on the fluid exerted by the particles should have the same
magnitude as the sum of the forces on all particles exerted by the fluid but with opposite
directions. 

{\color{black} The PCM-based coarse-graining schemes as in Eqs.~(\ref{eq:pcm-k}), (\ref{eq:pcm-u}),
  and (\ref{eq:pcm-f}) are conservative \emph{by construction}. Specifically, it conserves total
  particle mass, total particle momentum, and total momentum of the fluid--particle system. The
  proposed coarse-graining algorithm consists of two steps: (1) coarse graining using PCM, and (2)
  solving diffusion equations for the appropriate quantities.  The conservation requirements above
   dictate that diffusion equations should be solved for the following three quantities:
\begin{equation}
\varepsilon_s, \quad
 \varepsilon_s \mathbf{U}_s , \quad
\textrm{ and } \quad
 \varepsilon_f  \mathbf{F}^{fp} .
\end{equation}
The physical meaning of the three quantities above are particle mass, particle phase momentum, and
fluid--particle interaction forces per unit volume, respectively. Detailed derivations are presented
in the Appendix.  For vector fields such as $\varepsilon_s \mathbf{U}_s$ and $\varepsilon_f
\mathbf{F}^{fp}$, diffusion equations are solved for each component of the vector individually,
leading to seven diffusion equations in total for the three field variables. Conservation
requirements are thus met for all components.}

\subsubsection{Merits and Limitations of the Diffusion-Based Coarse-Graining Method}
The advantages of the diffusion-based coarse-graining method are extensively discussed and
demonstrated in~\cite{part1} via \emph{a priori} tests. The merits are summarized as follows:
\begin{enumerate}
\item sound theoretical foundation with equivalence to statistical kernel-based methods,
\item unified treatment of interior and near-boundary particles within the same framework,
\item guaranteed conservation of relevant physical quantities in the coarse-graining procedure,
\item easy implementation in CFD solvers with almost arbitrary meshes and ability to produce smooth
  and mesh-independent coarse-grained fields on unfavorable meshes, and
\item easy parallelization by utilizing the existing infrastructure in the CFD solver.
\end{enumerate}
{\color{black}
Potential limitations of the proposed method are summarized below:
\begin{enumerate}
\item Rigorous equivalence between the proposed method and statistical kernel-based methods only
  holds theoretically, i.e., when the CFD mesh is infinitely fine. Numerical diffusions can occur
  (compared with the results of the statistical kernel methods) on meshes with large cells,
  particularly when the diffusion bandwidth is smaller than the cell size. This shortcoming can be
  mitigated by setting the diffusion constant in the regions with large cells to very small values
  locally, effectively degenerating it to PCM in these regions.
\item The computational overhead associated with the diffusion-based method is significant, often
  exceeding the computational cost of the CFD solver.  However, in practical simulations where the
  number of particles per cell is reasonably large, the computational expense of the DEM solver
  dominates, and the overhead incurred by the coarse-graining procedure only accounts for a small
  fraction of the total computational cost.
\item Diffusing the fluid--particle drag forces in the proposed method makes it difficult to
  linearize and implicitly treat the fluid--particle momentum exchange terms in the fluid momentum
  equations. 
\end{enumerate}
Readers are referred to the companion paper~\citep{part1} and the rest of the present paper for
details.}

\subsection{Solver Implementation and Numerical Methods}
\label{sec:lammps-foam}
A hybrid CFD--DEM solver is developed based on two state-of-the-art open-source codes in their
respective fields, i.e., a CFD platform OpenFOAM (Open Field Operation and Manipulation) developed
by \citet{openfoam} and a molecular dynamics simulator LAMMPS (Large-scale Atomic/Molecular
Massively Parallel Simulator) developed at the Sandia National Laboratories~\citep{lammps}. This
hybrid solver was originally developed by the second author and his co-workers to study particle
segregation dynamics~\citep{sun09}. The solver was later used as a test bed for evaluating coarse
graining and sub-stepping algorithms in CFD--DEM~\citep{xiao-cicp}.  Recently, we have improved the
original solver significantly by enhancing its efficiency in the coupling of OpenFOAM and LAMMPS,
its parallel computing capabilities, and the coarse-graining algorithm, the last of which is the
subject of the current work.

The fluid equations in~(\ref{eq:NS}) are solved in OpenFOAM with the finite volume
method \citep{jasak96ea}. The solution algorithm is partly based on the work of~\cite{rusche03co} on
bubbly two-phase flows. The discretization is based on a collocated grid, i.e., pressure and all
velocity components are stored in cell centers. PISO (Pressure Implicit Splitting Operation)
algorithm is used to prevent velocity--pressure decoupling~\citep{issa86so}.  A second-order upwind
scheme is used for the spatial discretization of convection terms. A second-order central scheme is
used for the discretization of the diffusion terms.  Time integrations are performed with a
second-order implicit scheme.

The solution of the particle motions including their interactions via collisions and endured
contacts are handled by LAMMPS. The fluid forces \( \xmb{f}^{fp} \)  on the particles are computed
in OpenFOAM and supplied into LAMMPS and for its use in the integration of particle motion
equations~(\ref{eq:newton}). The particle forces on the fluid are computed in OpenFOAM according to
the forces on individual particles via a coarse-graining procedure.

The coarse-graining method used in this work involves solving transient diffusion equations.
Solution procedures of these equations are implemented based on the OpenFOAM platform, taking
advantage of existing infrastructure (e.g., discretization schemes, linear solvers, and parallel
computing capabilities) available in OpenFOAM. The diffusion equations are solved on the same mesh
as the CFD mesh. A second-order central scheme is used for the spatial discretization of the
diffusion equation; the Crank--Nicolson scheme is used for the temporal integration, which
guarantees the stability and allows for large time step sizes for the solution of the diffusions
equations to minimize computational overhead associated with the coarse graining.

\section{Numerical Simulations}
\label{sec:insitu}
In the companion paper~\citep{part1},  \emph{a priori} tests have been performed to highlight the
merits of the diffusion-based coarse-graining method by calculating the coarse-grained solid volume
field of a given particle configuration. The purpose of the present numerical tests is to examine
the performance of the new coarse-graining method in the context of a CFD--DEM solver applied to
fluidized bed flows.  

The CFD--DEM solver used in this study has been validated extensively by the second author and his
collaborators~\citep{sun09,xiao-cicp,gupta11a,gupta11b,gxs11a,gupta12,gupta13a}, some of which were
conducted within an EU-funded project PARDEM~\citep{pardem}. Here we present only a brief validation
of the current solver with CFD--DEM simulations in the literature based on the same experimental
setup used in this work. Then, the mesh-convergence tests are performed on the CFD--DEM solver with
the diffusion-based coarse-graining method. This is a follow-up investigation of the a priori
mesh-convergence tests presented in the companion paper~\citep{part1}. The purpose of this test is to
demonstrate the capability of the diffusion-based coarse-graining algorithm in yielding
mesh-converged results in CFD--DEM simulation, which has been a major challenge so far, particularly
when the cell sizes are small compared to the particles. Finally, numerical tests are performed by
using the CFD--DEM solvers with the diffusion-based coarse graining and with the PCM-based coarse
graining. This is to highlight the advantages of the diffusion-based coarse graining both in
terms of producing mesh-independent results and in representing correct physical mechanisms in 
dense particle-laden flows.

\begin{figure}[!htpb]
  \centering
  \includegraphics[width=0.55\textwidth]{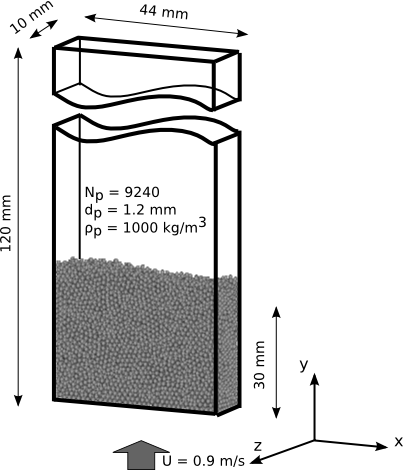}
  \caption{Geometry of the 3D computational domain for the simulation.}
  \label{fig:insitu-geometry}
\end{figure}

The numerical tests are set up based on the fluidized bed experiments of
\citep{muller08gr,muller09va}. In the experiments, the dimensions of the fluidized bed were $44$~mm
$\times~1500$~mm $\times~10$~mm (width, height, and transverse thickness, aligned with the $x$, $y$,
and $z$ axes, respectively, in our coordinate system).  The domain geometry is shown in
Fig.~\ref{fig:insitu-geometry} along with the coordinate system used here. In their experiments,
magnetic resonance was used to measure the volume fraction $\varepsilon_f$ of the fluid (i.e.,
air)~\citep{muller09va} and the velocity of particles~\citep{muller08gr}. The superficial inlet
velocity of the air is $0.9$~m/s, and the initial bed is approximately $30$~mm in height, consisting
of 9240 poppy seed particles. Other parameters of the experiments are summarized in
Table~\ref{tab:insitu-exp}. To reduce the computational costs without comprising the accuracy of the
numerical simulations, the height of the bed is taken as $120$~mm following~\cite{muller08gr}.
Although the poppy seeds are kidney shaped and thus are not exactly spherical particles, for
simplicity they are considered spherical in the simulations here. Slip conditions are applied at the
boundaries in the $z$-direction, and no-slip boundary conditions are applied in the $y$-direction
boundaries. The time step in the DEM simulations is taken as $4.0 \times 10^{-6}$~s. To get the
time-averaged profiles of fluid volume fraction and particle velocity, the simulations are averaged
for 18~s, which is approximately 135 flow-through times, and is long enough to achieve statistically
converged time-averaging fields~\citep{muller09va}.  The choices of computational setup and
parameters outlined above are consistent with previous numerical validations of this set of
experiments~\citep{muller08gr,muller09va}. The bandwidth $b$ used in the diffusion-based
coarse-graining method is $4d_p$ (with $d_p$ being the particle diameter).

\begin{table}[!htbp]
  \caption{Parameters used in the CFD--DEM simulations of the fluidized bed flow.}
 \begin{center}
 \begin{tabular}{ll}
   \hline
   bed dimensions & \\
   \qquad width $(L_x)$ & 44~mm\\
   \qquad height $(L_y)$ & 120~mm\\
   \qquad transverse thickness $(L_z)$ & 10~mm\\
   particle properties & \\
   \qquad total number & $9240$\\
   \qquad diameter $d_p$ & 1.2~mm\\
   \qquad density $\rho_s$ & $1.0 \times 10^3~\mathrm{kg/m^3}$\\
   \qquad elastic modulus & $1.2 \times 10^{-5}$~Pa\\
   \qquad Poisson's ratio & 0.33\\
   \qquad normal restitution coefficient & 0.98\\
   \qquad coefficient of friction & 0.1\\
   fluid properties & \\
   \qquad density $\rho_f$ & $1.2~\mathrm{kg/m^3}$\\
   \qquad viscosity & $1.8 \times 10^{-5}~\mathrm{kg/(m \cdot s)}$\\
   \qquad superficial inlet velocity & $0.9$~m/s\\
   \hline
  \end{tabular}
 \end{center}
 \label{tab:insitu-exp}
\end{table}

\subsection{Solver Validations}
\label{sec:insitu-valid}

The purpose of this validation is to show the results obtained from the proposed CFD--DEM solver are
consistent with the experimental measurements and numerical simulations in the literature. The
setup and parameters used in this validation test are detailed in Table~\ref{tab:insitu-exp}. The
mesh resolution used in this simulation is $N_x \times N_y \times N_z = 36 \times 100 \times 8$,
where $N_x$, $N_y$, and $N_z$ are the numbers of cells in the width ($x$-), height ($y$-), and
transverse thickness ($z$-) directions, respectively.

\begin{figure}[!htpb]
  \centering
  \subfloat[][$y=16.4$~mm]{
    \label{fig:voidageValidation1}
    \includegraphics[width=0.45\textwidth]{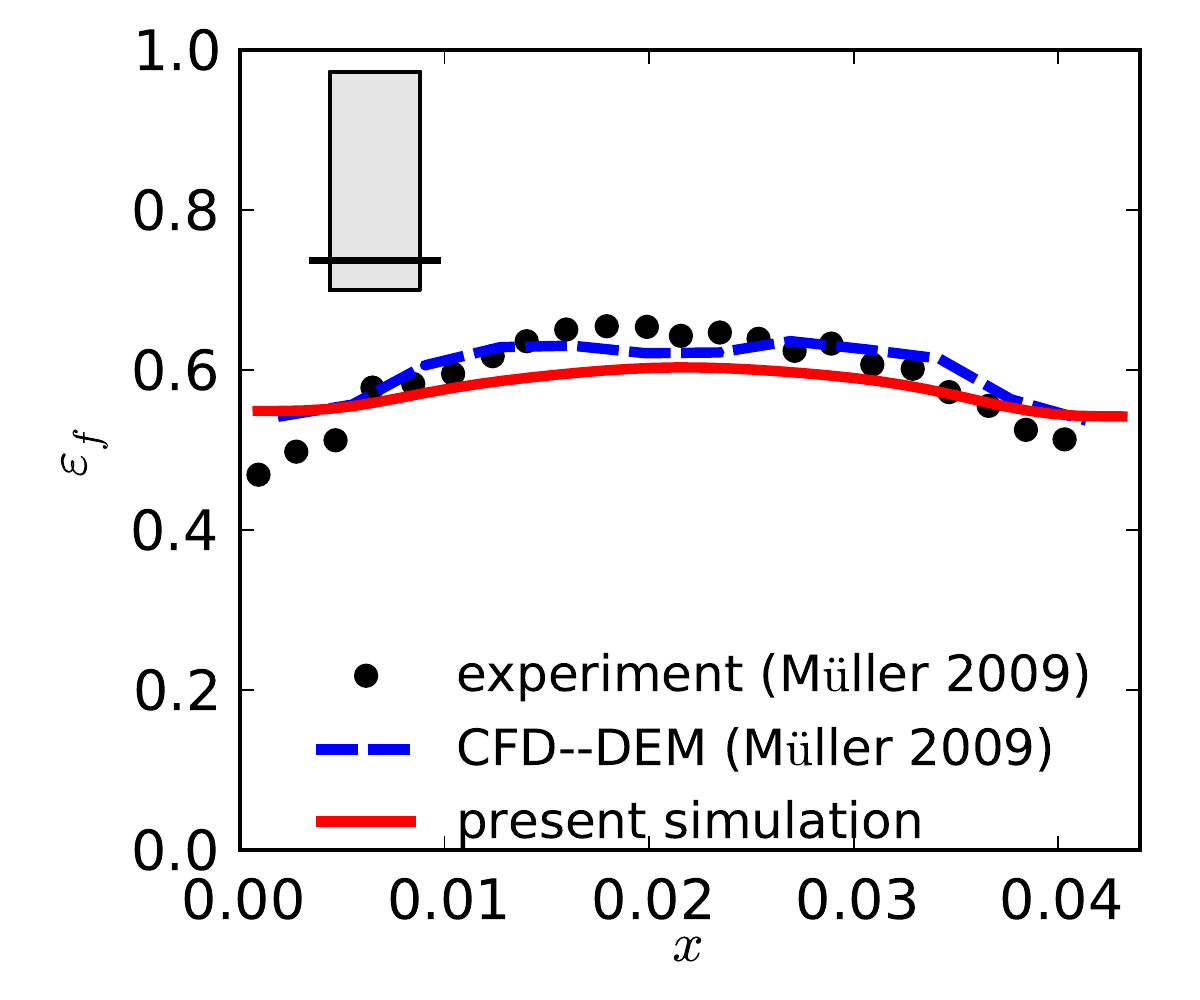}
  }
  \hspace{0.01\textwidth}
  \subfloat[][$y=31.2$~mm]{
    \label{fig:voidageValidation2}
    \includegraphics[width=0.45\textwidth]{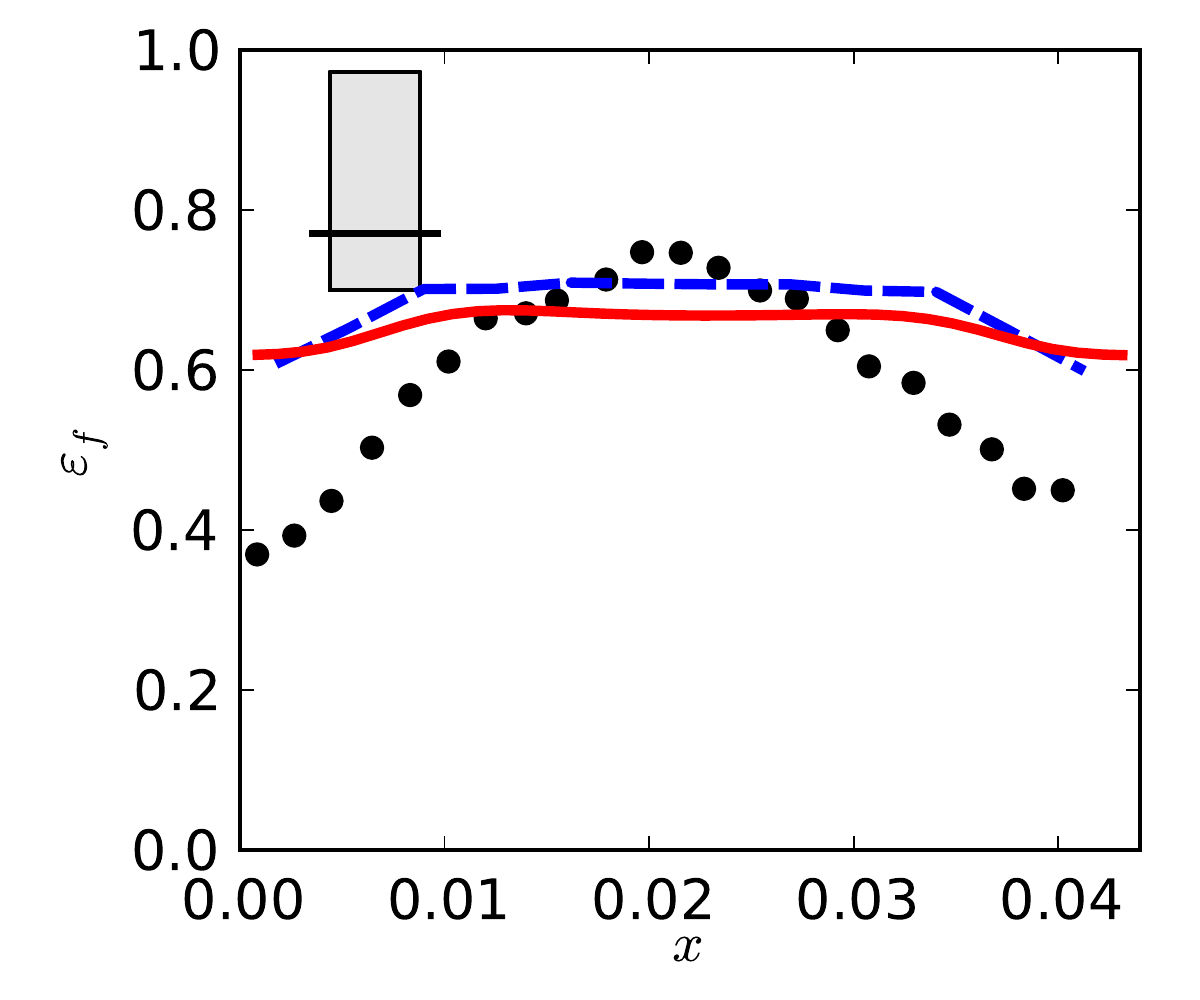}
  }
  \caption{Profiles of time-averaged fluid volume fraction obtained from the present simulations
    compared with the experimental measurements and numerical simulation of~\citet{muller09va}. The
  results are presented at two vertical locations: (a) $y = 16.4$~mm and (b) $y= 31.2$~mm.}
  \label{fig:voidageValidation}
\end{figure}

The measurements of fluid volume fraction $\varepsilon_f$ in the experiments~\citep{muller09va} were
taken on two cross-sections at the heights $y = 16.4$~mm and $y = 31.2$~mm. To facilitate comparison
with the experimental measurements, the profiles of $\varepsilon_f$ at the two heights are extracted
in the present simulations. The comparison with the experimental data and the CFD--DEM simulations
of \citet{muller09va} are presented in Fig.~\ref{fig:voidageValidation}. As can be seen from the
figure, overall the $\varepsilon_f$ profiles predicted by the present CFD--DEM solver have favorable
agreement with the experimental results and the numerical simulations of
\citet{muller09va}. However, it is noted that both solvers predicted higher fluid volume fraction
$\varepsilon_f$ near the boundaries (slippery walls) compared with experimental measurements. This
is particularly prominent for location $y = 31.2$~mm.  As pointed out by \citet{muller09va}, the
discrepancy in the $\varepsilon_f$ profiles is due to the over-prediction of the width of the
bubbles in the CFD--DEM solvers, which consequently leads to higher $\varepsilon_f$ values near the
wall.  Since this over-prediction is attributed to the difficulty of the CFD--DEM framework in
modeling the wall effects, and it is not directly related to our solver in particular, further
discussions are not pursued here.

\begin{figure}[!htpb]
  \centering
  \subfloat[][$h=15$~mm]{
    \label{fig:velocityValidation1}
    \includegraphics[width=0.45\textwidth]{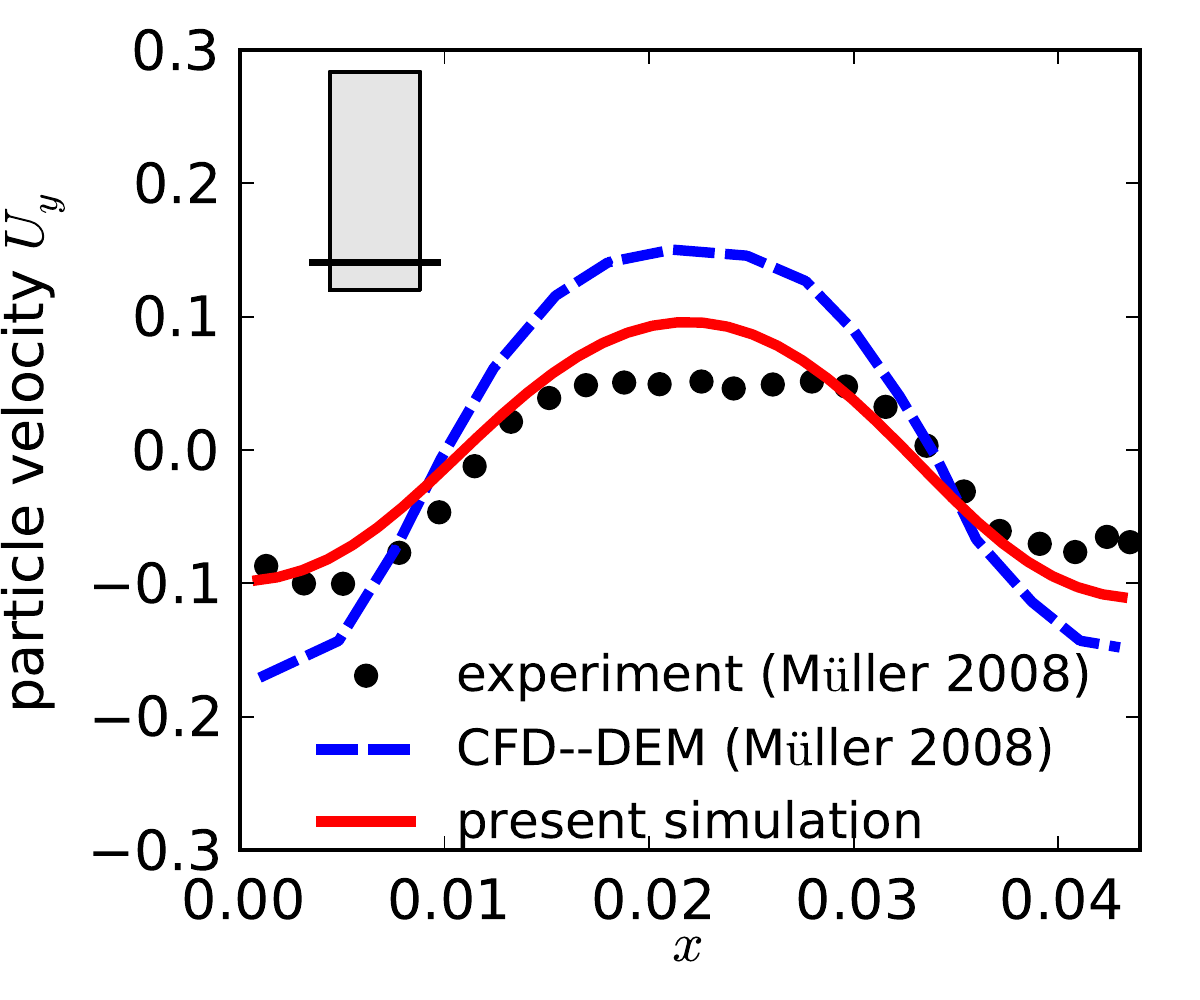}
  }
  \hspace{0.01\textwidth}
  \subfloat[][$h=25$~mm]{
    \label{fig:velocityValidation2}
    \includegraphics[width=0.45\textwidth]{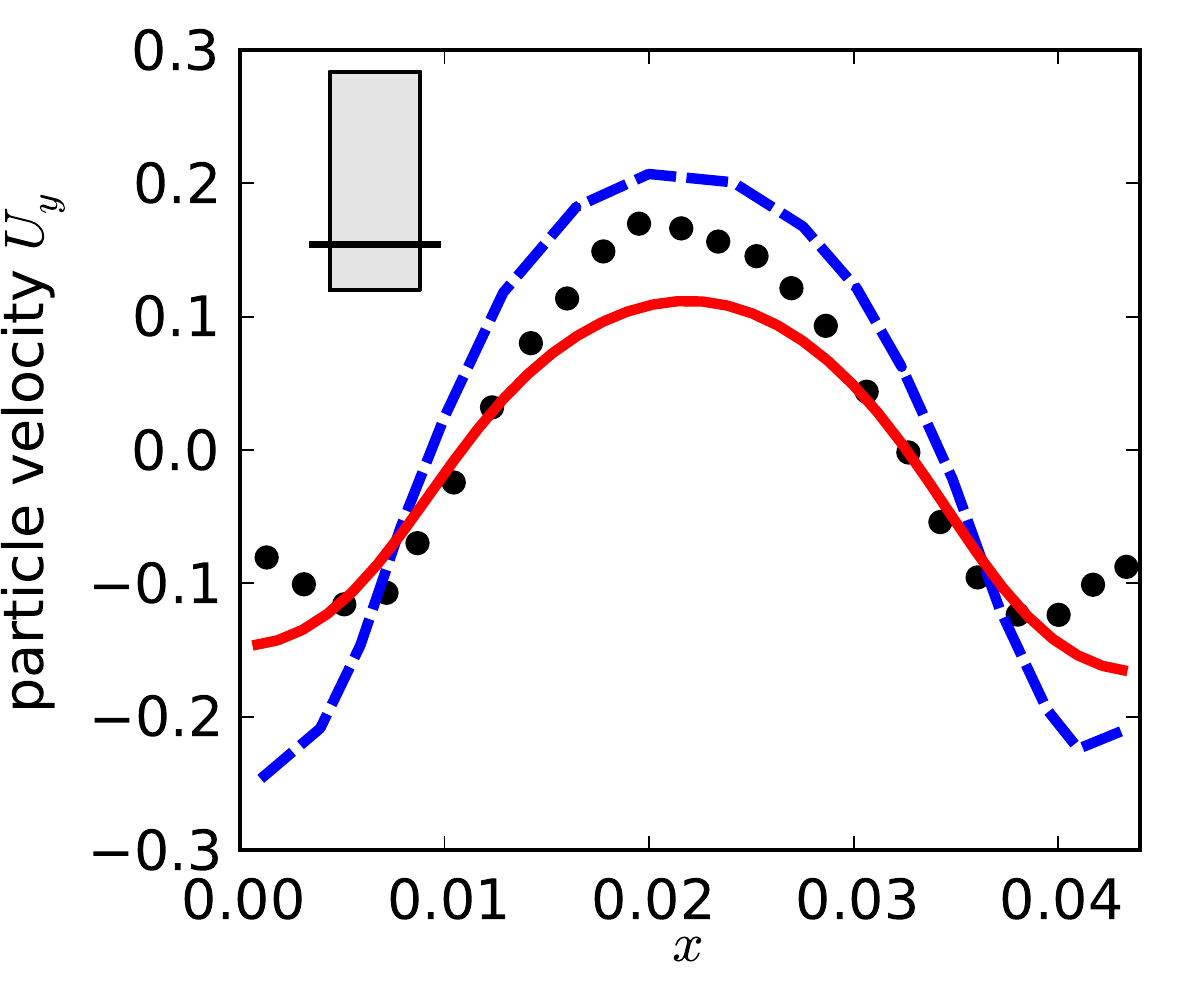}
  }
  \caption{Profiles of time-averaged vertical velocity of particles obtained from the present
    simulation. The measurements are taken at different heights according to the
    experiment~\citep{muller08gr}. (a) Particle velocity at the height of 15~mm, (b) particle
    velocity at the height of 25~mm.}
  \label{fig:velocityValidation}
\end{figure}

The vertical component $U_y$ of the time-averaged particle phase velocity are shown in
Fig.~\ref{fig:velocityValidation}, with comparisons among the current simulations, the experimental
measurement, and the numerical simulations of \citeauthor{muller08gr} The comparisons are presented for
two cross-sections at the heights $y = 15$~mm and $y=25$~mm, as these are the locations where
experimental measurements were performed~\citep{muller08gr}. It can be seen from
Fig.~\ref{fig:velocityValidation} that generally speaking both CFD--DEM solvers give good
predictions of the particle phase velocity $U_y$ for both locations, although arguably the
prediction quality of our solver seems to be slightly better.  Specifically, the simulations of
\citeauthor{muller08gr} tend to over-predict the particle velocities at both locations; while our
simulations do not seem to have this issue except for a very minor over-prediction near the
centerline (between $x = 0.015$~m and $0.025$~m) for the profile at $y = 15$~mm. On the other hand,
the particle velocity $U_y$ near the centerline is under-predicted for $y = 25$~mm.  Overall, the
magnitude of time-averaged particle velocity predicted by the present simulation using
diffusion-based algorithm is smaller than that in the simulation of \citeauthor{muller08gr} A possible
explanation is the different solid volume fraction $\varepsilon_s$ fields used in the two
solvers. The $\varepsilon_s$ field computed with the diffusion-based method is smoother and is thus
free from very large $\varepsilon_s$ values. Since the computed drag force on a particle increases
dramatically with the increase of $\varepsilon_s$, over-prediction of the drag forces on particles
due to high $\varepsilon_s$ is more likely when the PCM coarse graining is used compared to the
diffusion-based method.  It is worth noting that one should not be deceived by the very similar
averaged $\varepsilon_s$ profiles for both simulations in Fig.~\ref{fig:voidageValidation}. In fact,
the instantaneous $\varepsilon_s$ fields obtained with PCM have much more very larger values, as has
been demonstrated in the \emph{a priori} tests in \citet{part1}. Although present only in very few
cells, these large values can have a significant impact on the particle velocities, since the drag
force is a highly nonlinear function of $\varepsilon_s$.  Consequently, this leads to larger fluid
drag forces on the particles in the PCM-based CFD--DEM solvers, and thus larger averaged solid phase
velocities in the results.

In summary, although some discrepancies exist between the prediction of $\varepsilon_f$ by the
present CFD--DEM solver and the experimental results, the overall agreement is favorable.  Moreover,
the fluid volume fraction and particle phase velocities obtained in the present simulations are in
very good agreement with the previous CFD--DEM simulations of the same case~\citep{muller08gr}.
Hence, the validations are deemed successful, and further investigations using the CFD--DEM solver
are pursued below.

\subsection{Mesh-Independence Study of CFD--DEM simulations}
\label{sec:insitu-indep}

In the \emph{a priori} simulations presented in the companion paper~\citep{part1}, it has been
demonstrated that the diffusion-based method yields mesh-independent solid volume fraction
fields. Here we further demonstrate that the CFD--DEM simulations with diffusion-based coarse
graining give mesh-independent results. In particular, average particle phase velocities and fluid
volume fraction at various locations are studied extensively and are compared among simulations
performed on five successively refined meshes.  To characterize the sizes of arbitrarily shaped
cells, we use the parameter \emph{cell length scale} $S_c$ defined as:
\begin{equation}
  S_c = \sqrt[3]{V_c} ,
  \label{eqn:insitu-size}
\end{equation}
where $V_c$ is the volume of a cell.  Five meshes A--E are used in the mesh-independence studies
here with the effective lengths $S_c$ = $6d_p$ (mesh A), $4d_p$ (mesh B), $2d_p$ (mesh C), $d_p$
(mesh D), and $0.5d_p$ (mesh E), arranged with increasing mesh resolution. Larger $S_c$ values
indicate larger cell sizes and thus coarser meshes. Details of the mesh parameters are presented in
Table~\ref{tab:insitu-indep}.

\begin{table}[!htbp]
  \caption{
  \label{tab:insitu-indep}
  Parameters in mesh-independence study of the CFD--DEM solvers with diffusion-based and PCM-based coarse-graining
  methods. The study using the solver with diffusion-based method covers a wider range of mesh
  resolutions, while meshes (e.g., D and E) with small cells cannot be used in the PCM-based solver
  due to instabilities.}
  \begin{center}
  \begin{tabular}{l|ccccc}
    \hline
    mesh & $S_c/d_p$ & $N_x$ & $N_y$ & $N_z$ & used in \\
    \hline
    A & 6 & $6$ & $16$ & $1$ & diffusion-based; PCM \\
    B & 4 & $9$ & $25$ & $2$ & diffusion-based; PCM \\
    B$^{\prime}$ & 3 & $12$ & $32$ & $3$ & PCM only \\
    C & 2 & $18$ & $50$ & $4$ & diffusion-based; PCM \\
    D & 1 & $36$ & $100$ & $8$ & diffusion-based only \\
    E & 0.5 & $72$ & $200$ & $16$ & diffusion-based only \\
    \hline
  \end{tabular}
  \end{center}
\end{table}

\begin{figure}[!htpb]
  \centering
  \subfloat[][$y=16.4$~mm]{
    \includegraphics[width=0.45\textwidth]{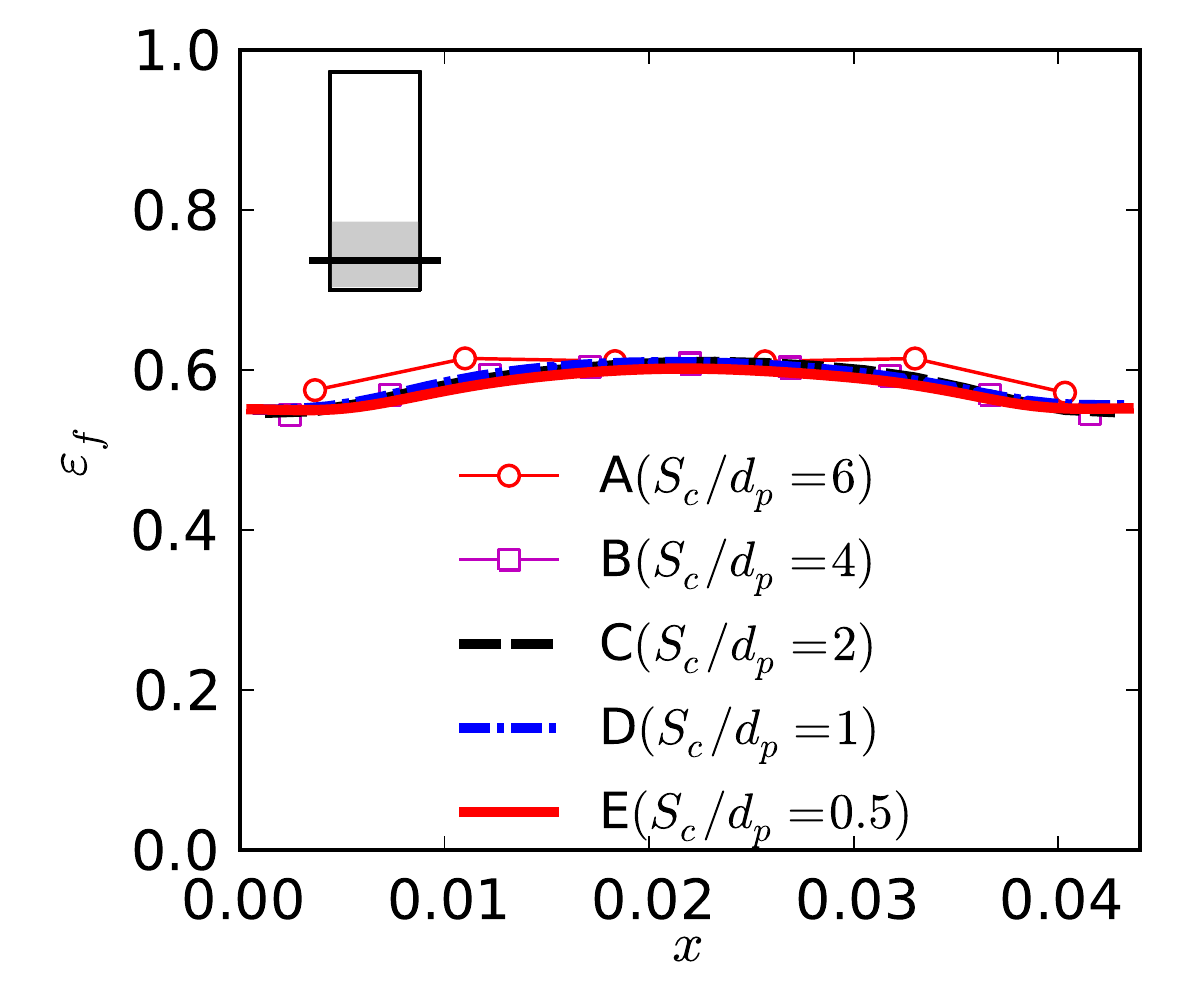}
  }
  \hspace{0.01\textwidth}
  \subfloat[][$y=31.2$~mm]{
    \includegraphics[width=0.45\textwidth]{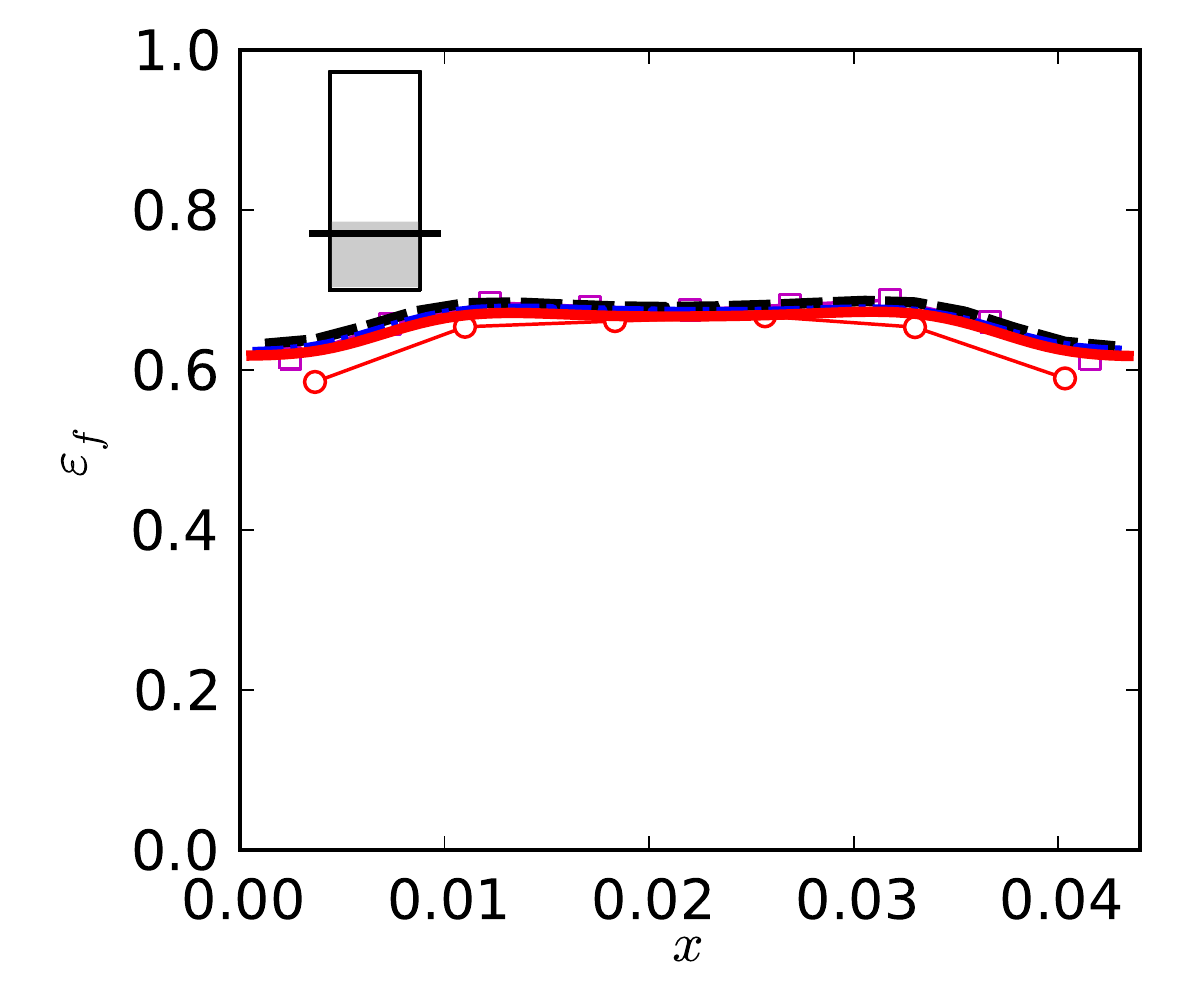}
  }
  \vspace{0.01\textwidth}
  \subfloat[][$x=11$~mm]{
    \includegraphics[width=0.45\textwidth]{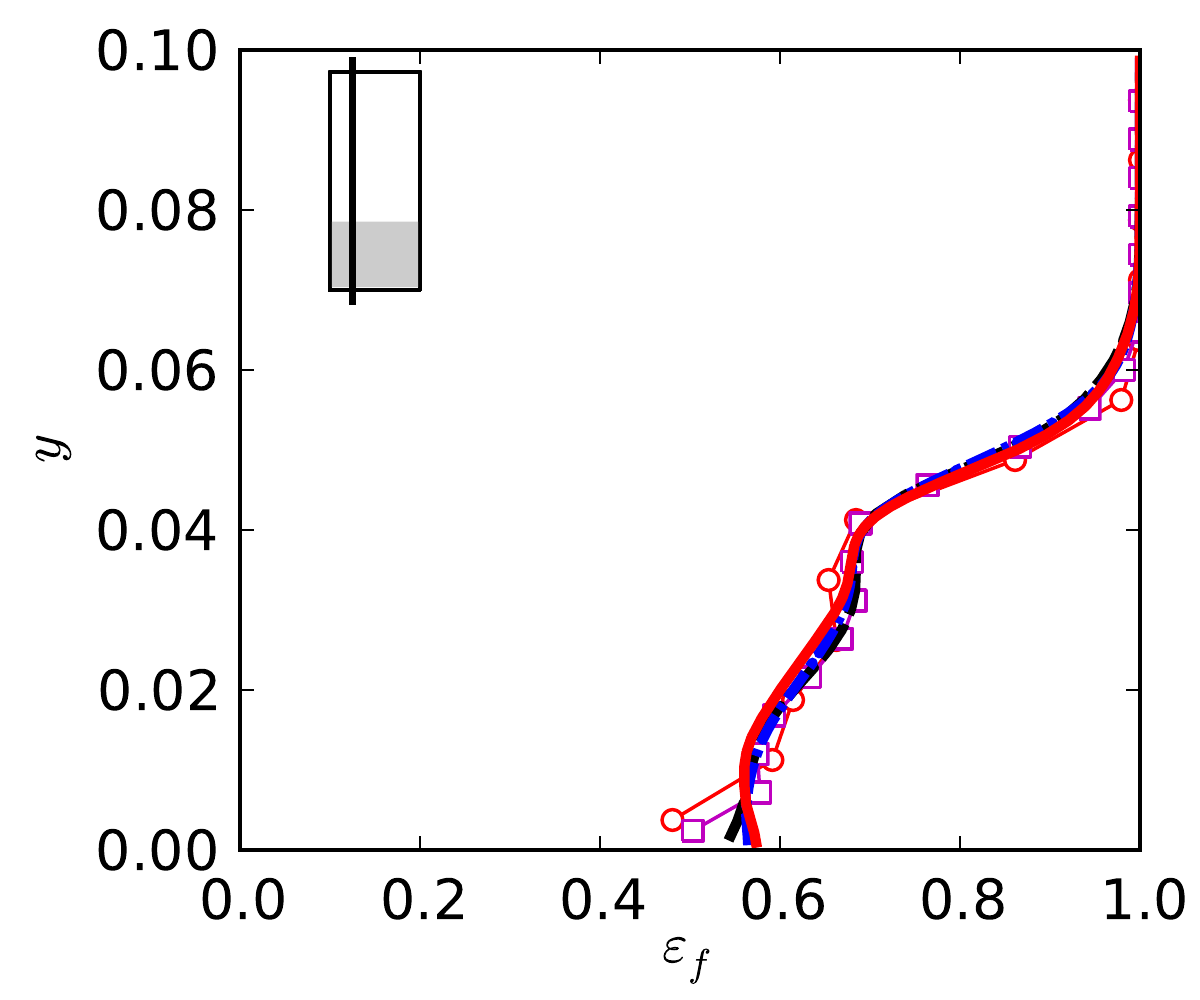}
  }
  \hspace{0.01\textwidth}
  \subfloat[][$x=22$~mm]{
    \includegraphics[width=0.45\textwidth]{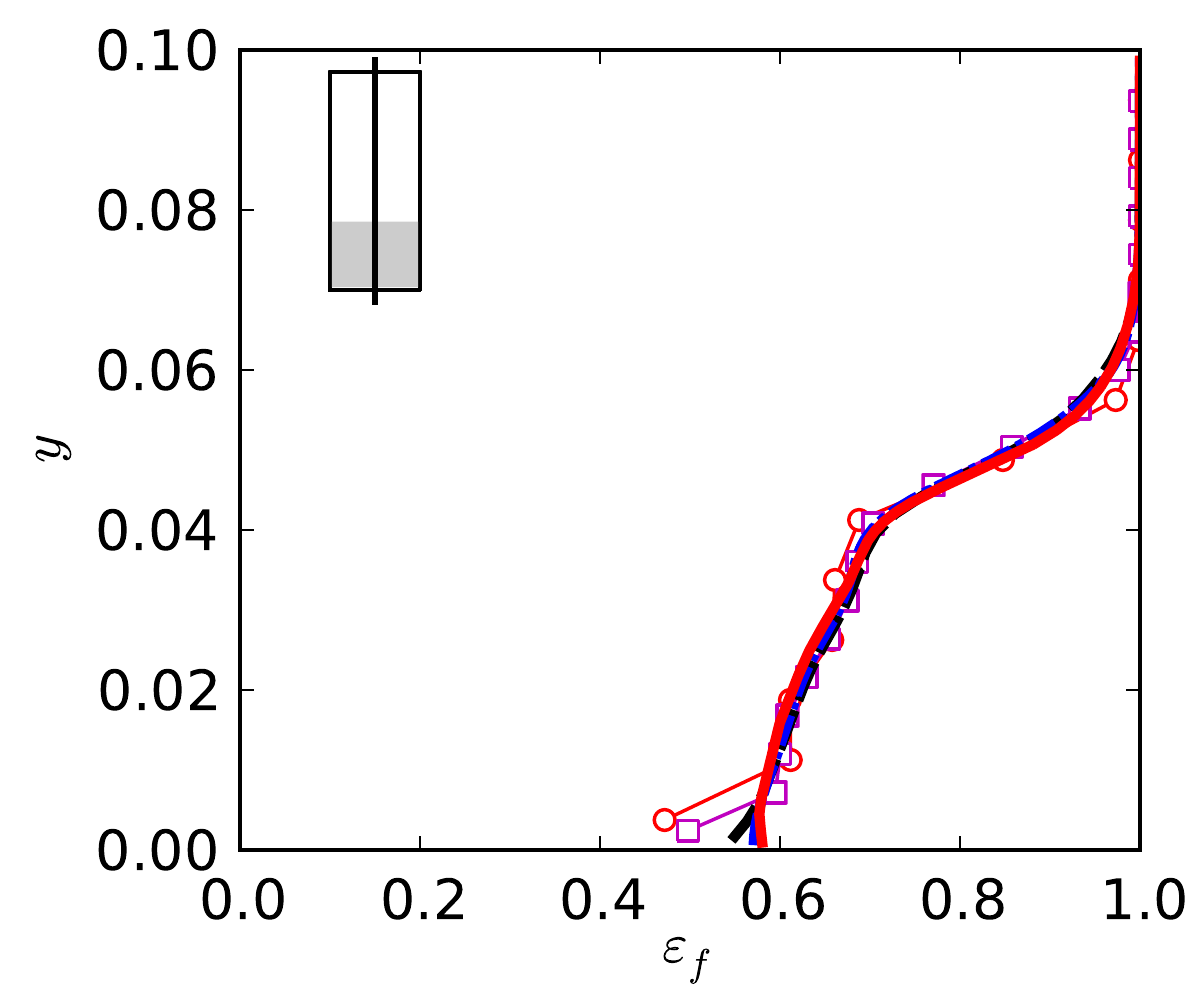}
  }
  \caption{Mesh convergence study of the CFD--DEM solver with the diffusion-based coarse-graining
    method, showing the fluid volume fraction $\varepsilon_f$ ($=1-\varepsilon_s$) profiles on two
    horizontal cross-sections located at (a) $y=16.4$~mm and (b) $y=31.2$~mm, respectively, and two
    vertical cross-sections located at (c) $x=11$~mm and (d) $x=22$~mm, respectively. Insets in the
    panels show the location of the cross-section corresponding to each profile.  The shaded regions
    in the inset indicates the initial particle bed.  Results obtained based on five consecutively
    refined meshes are compared.}
  \label{fig:voidageIndependence}
\end{figure}

As with the validation studies above, the fluid volume fractions $\varepsilon_f$ profiles at two
cross sections at heights $y=16.4$~mm and $y=31.2$~mm are shown in
Figs.~\ref{fig:voidageIndependence}(a) and \ref{fig:voidageIndependence}(b). Moreover, the profiles
of fluid volume fraction $\varepsilon_f$ at two vertical cross-sections at $x=11$~mm and $x=22$~mm
(located at a quarter width location and at the centerline of the domain, respectively) are shown in
Figs.~\ref{fig:voidageIndependence}(b) and \ref{fig:voidageIndependence}(d).  Although experimental
data are not available at the two vertical cross-sections, this does not impair the objective of the
mesh-convergence study, since we are mainly concerned with the comparison of results obtained with
meshes with different coarseness levels, and not with the agreement between numerical predictions
and experimental measurements.  Examining the profiles at a few vertical cross-sections in addition
to the horizontal cross-section locations helps shed light on the behavior of the result in the
entire domain. To facilitate visualization, the locations of the corresponding cross-sections are
indicated in the insets in the upper left corner of each plot along with the $\varepsilon_f$
profiles. The shaded regions in the insets indicate the initial particle bed.  It can be seen that
mesh convergence is achieved in the prediction of $\varepsilon_f$ as all meshes with $S_c/d_p$
smaller than 4 (e.g., meshes B, C, D, and E) give identical results. As to the $\varepsilon_f$
profiles obtained by using mesh A with $S_c/d_p=6$, some minor discrepancies with the converged
results are observed, particularly in the region near the bottom inlet in
Figs.~\ref{fig:voidageIndependence}(c) and ~\ref{fig:voidageIndependence}(d). Since this mesh
resolution is probably inadequate, the minor discrepancies are expected.

\begin{figure}[!htpb]
  \centering
  \subfloat[][$y=15$~mm]{
    \includegraphics[width=0.45\textwidth]{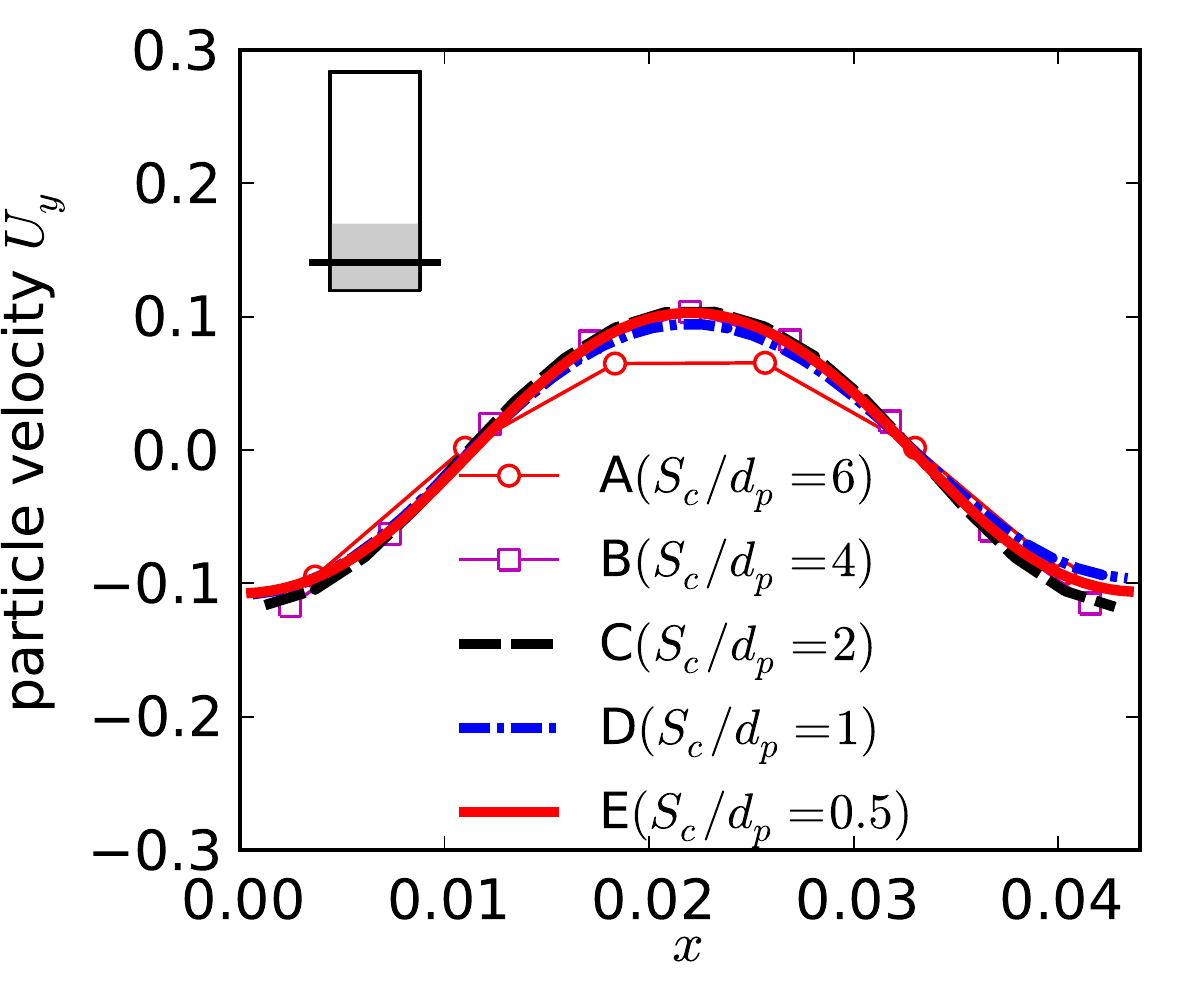}
  }
  \hspace{0.01\textwidth}
  \subfloat[][$y=25$~mm]{
    \includegraphics[width=0.45\textwidth]{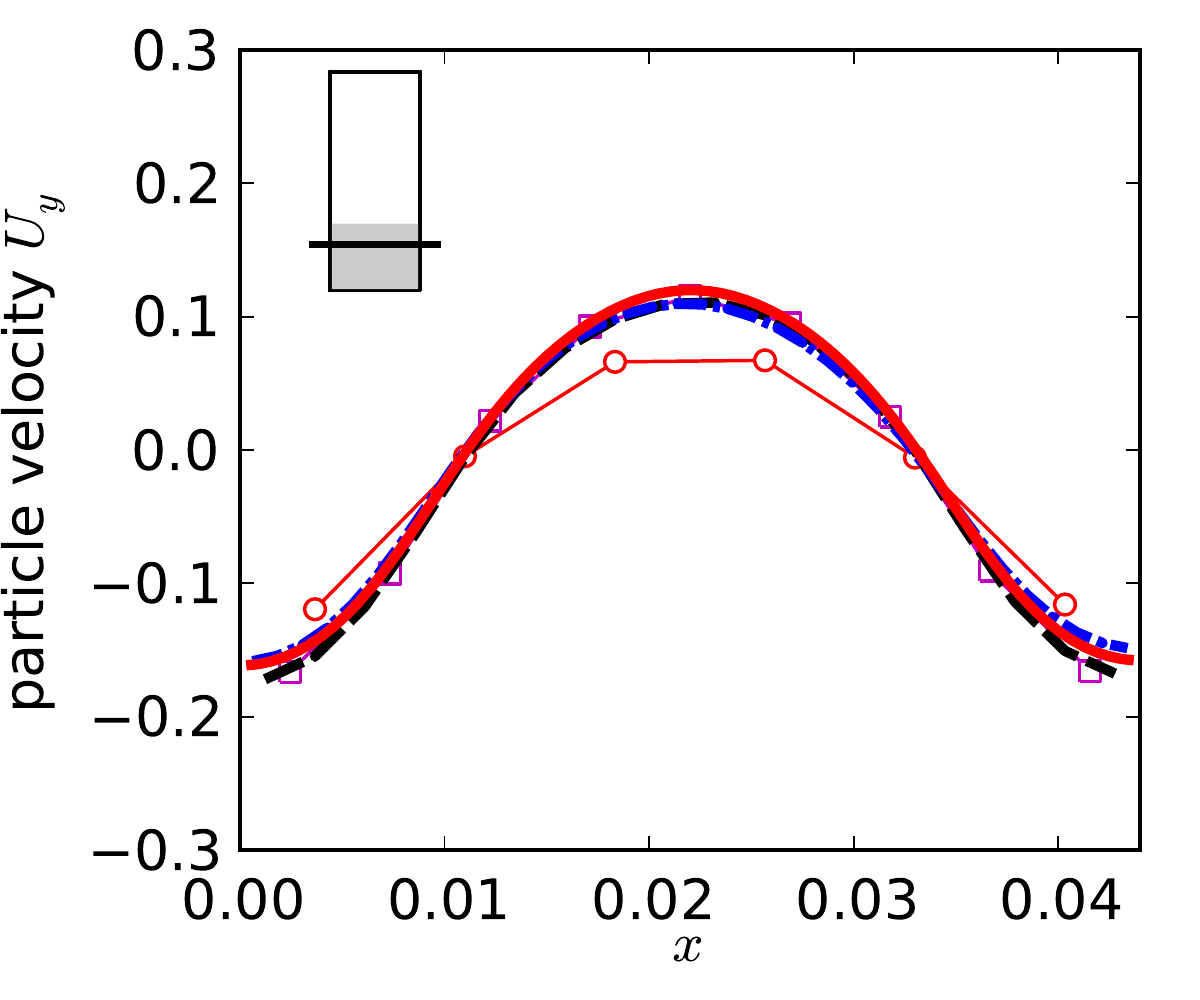}
  }
  \vspace{0.01\textwidth}
  \subfloat[][$y=11$~mm]{
    \includegraphics[width=0.45\textwidth]{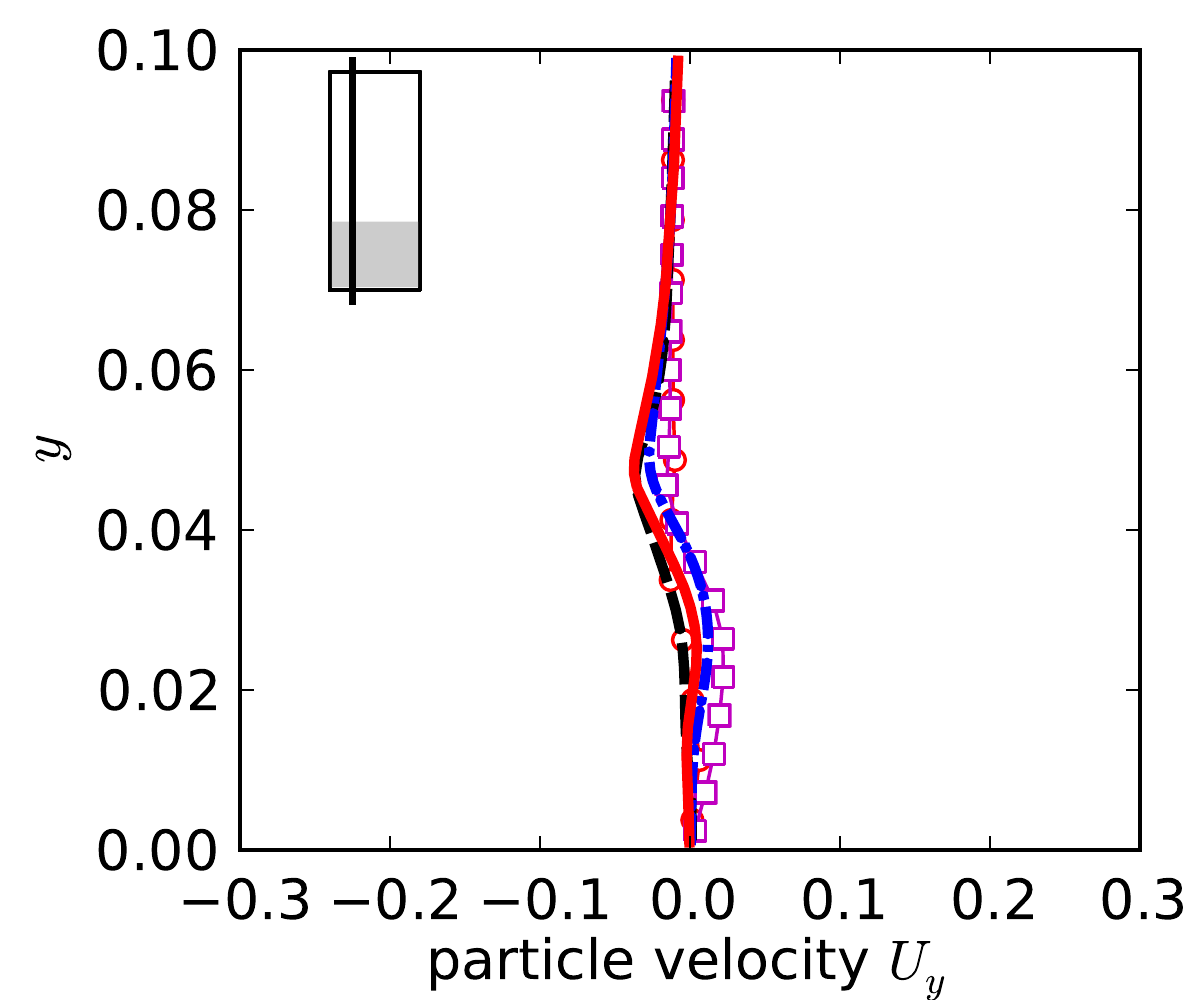}
  }
  \hspace{0.01\textwidth}
  \subfloat[][$y=22$~mm]{
    \includegraphics[width=0.45\textwidth]{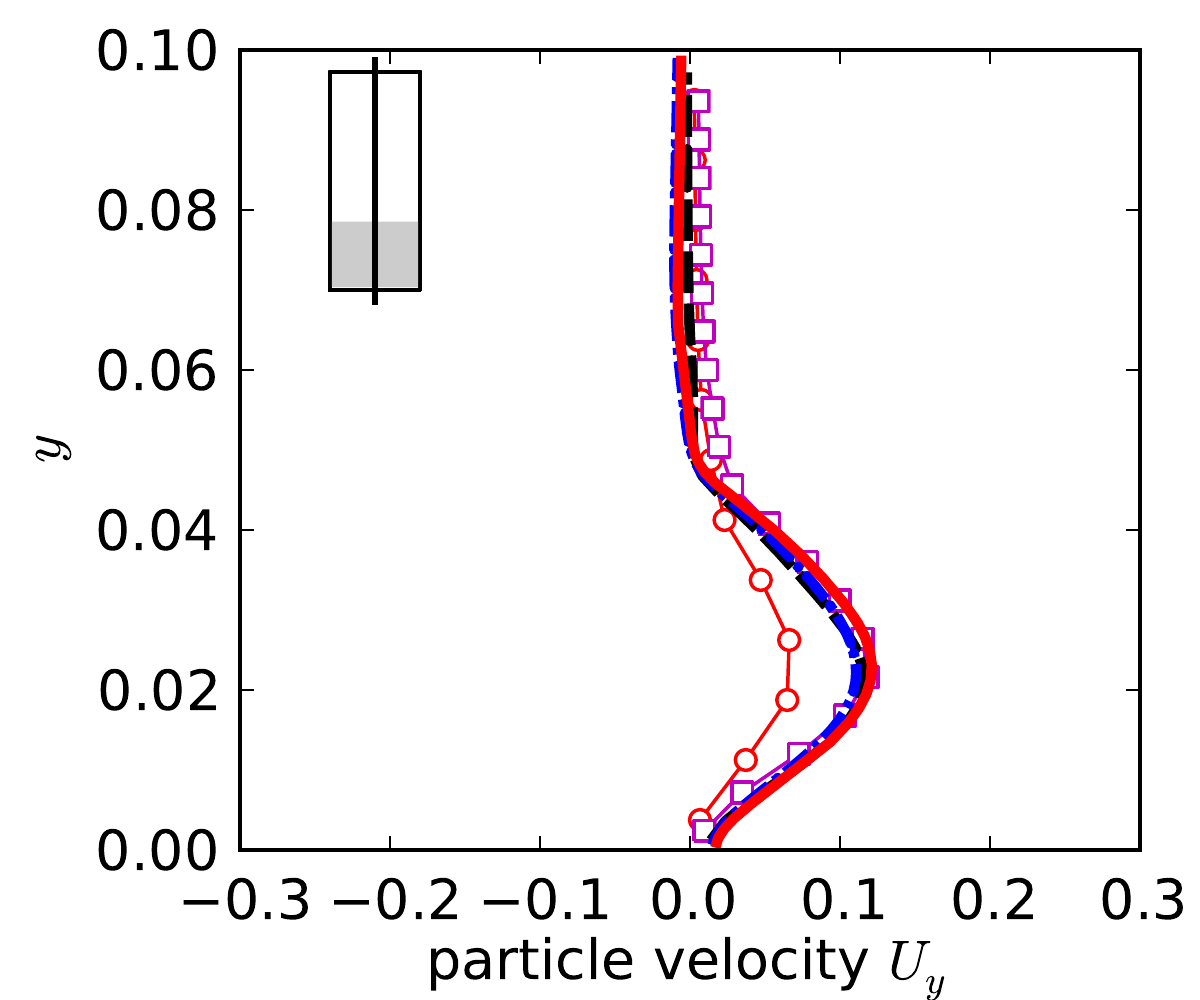}
  }
  \caption{Mesh convergence study of the CFD--DEM solver with the diffusion-based coarse-graining
    method, showing the profiles of vertical component $U_y$ of the time-averaged particle velocity
    obtained on two horizontal cross-sections located at (a) $y=16.4$~mm and (b) $y=31.2$~mm,
    respectively, and two vertical cross-sections located at (c) $x=11$~mm and (d) $x=22$~mm,
    respectively. Insets in the panels show the locations of the cross-section corresponding to each
    profile.  Results obtained based on five consecutively refined meshes are compared.}
  \label{fig:velocityIndependence}
\end{figure}

Similarly, the vertical component $U_y$ of the time-averaged particle phase velocity obtained at
different horizontal and vertical cross-sections are shown in
Fig.~\ref{fig:velocityIndependence}. Again, the general observation here is that the four finer
meshes (B--D) all give identical results, indicating excellent mesh-convergence behavior, while some
discrepancies are found in the results from mesh A ($S_c/d_p = 6$). In contrast to the fluid volume
fraction profiles shown in Fig.~\ref{fig:voidageIndependence}, the discrepancies between the results
from mesh A and the converged results from meshes B--E occur mostly in the middle of the domain
(e.g., between $x=0.01$ and $x=0.03$ in Figs.~\ref{fig:velocityIndependence}(a) and (b), and around
$y = 0.25$ in Figs.~\ref{fig:velocityIndependence}(c) and (d)).

\subsection{Performance Comparison with PCM}
\label{sec:insitu-perform}
The mesh-convergence studies above demonstrated excellent mesh-convergence behavior of the CFD--DEM
solver with diffusion-based coarse-graining method.  To highlight this advantage, the same
mesh-convergence study is performed on the CFD--DEM solver with PCM-based coarse-graining.  All
simulation setup and parameters are kept the same except for the coarse-graining method and the
meshes used.  Here only four meshes A ($S_c/d_p = 6$), B ($S_c/d_p = 4$), B$^\prime$ ($S_c/d_p =
3$), and C ($S_c/d_p = 2$) are used in the mesh-convergence study with PCM, since it is not possible
obtain meaningful solid volume fraction field on meshes D ($S_c/d_p = 1$) and E ($S_c/d_p = 0.5$),
which have cell length scales $S_c$ equal or smaller than the particle diameter. Even by capping the
$\varepsilon_s$ field at 0.7 (detailed below), we were not able to complete simulations on meshes D
or E without being interrupted by instabilities.

\begin{figure}[!htpb]
  \centering
  \subfloat[][$y=16.4$~mm]{
    \includegraphics[width=0.45\textwidth]{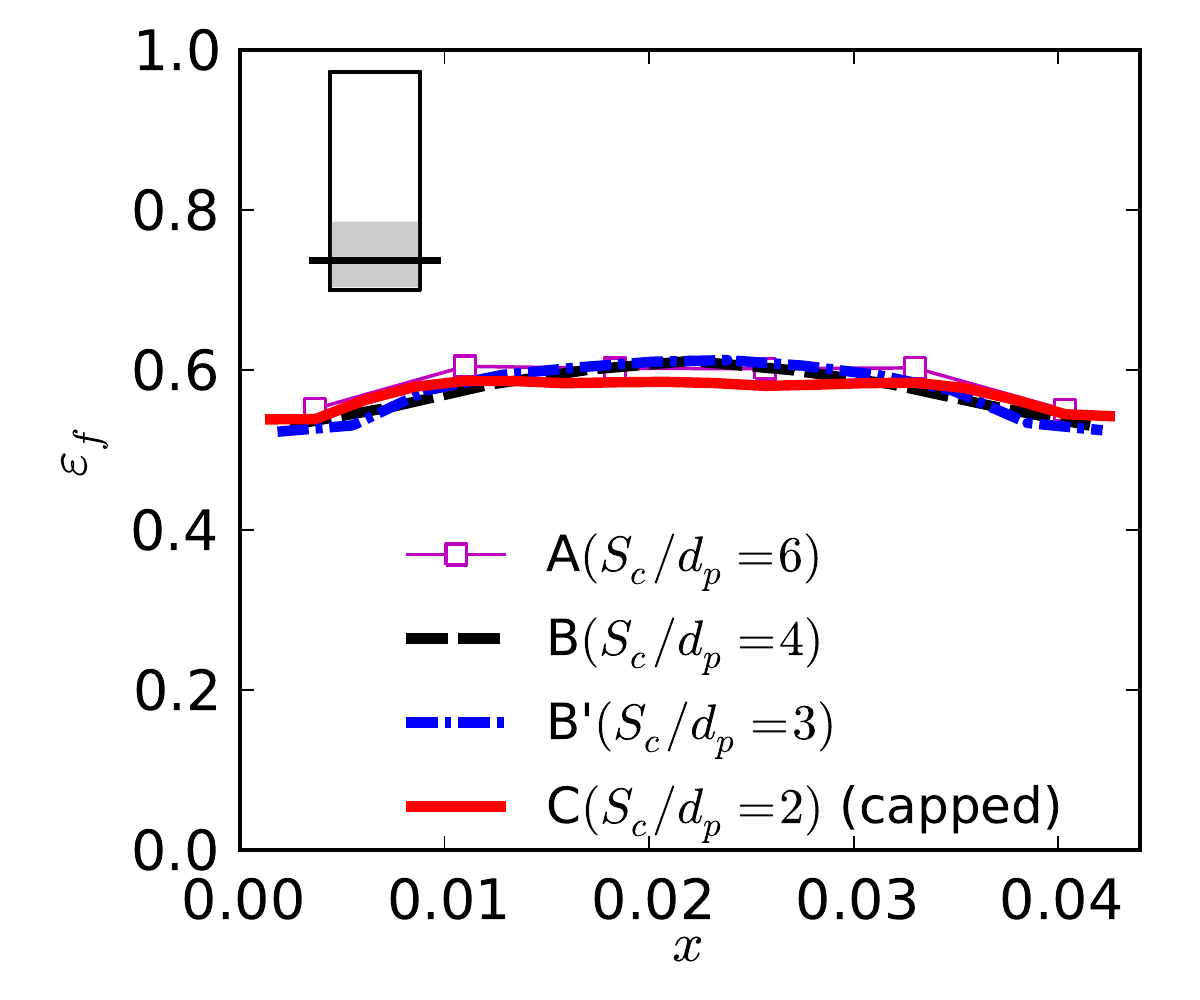}
  }
  \hspace{0.01\textwidth}
  \subfloat[][$y=31.2$~mm]{
    \includegraphics[width=0.45\textwidth]{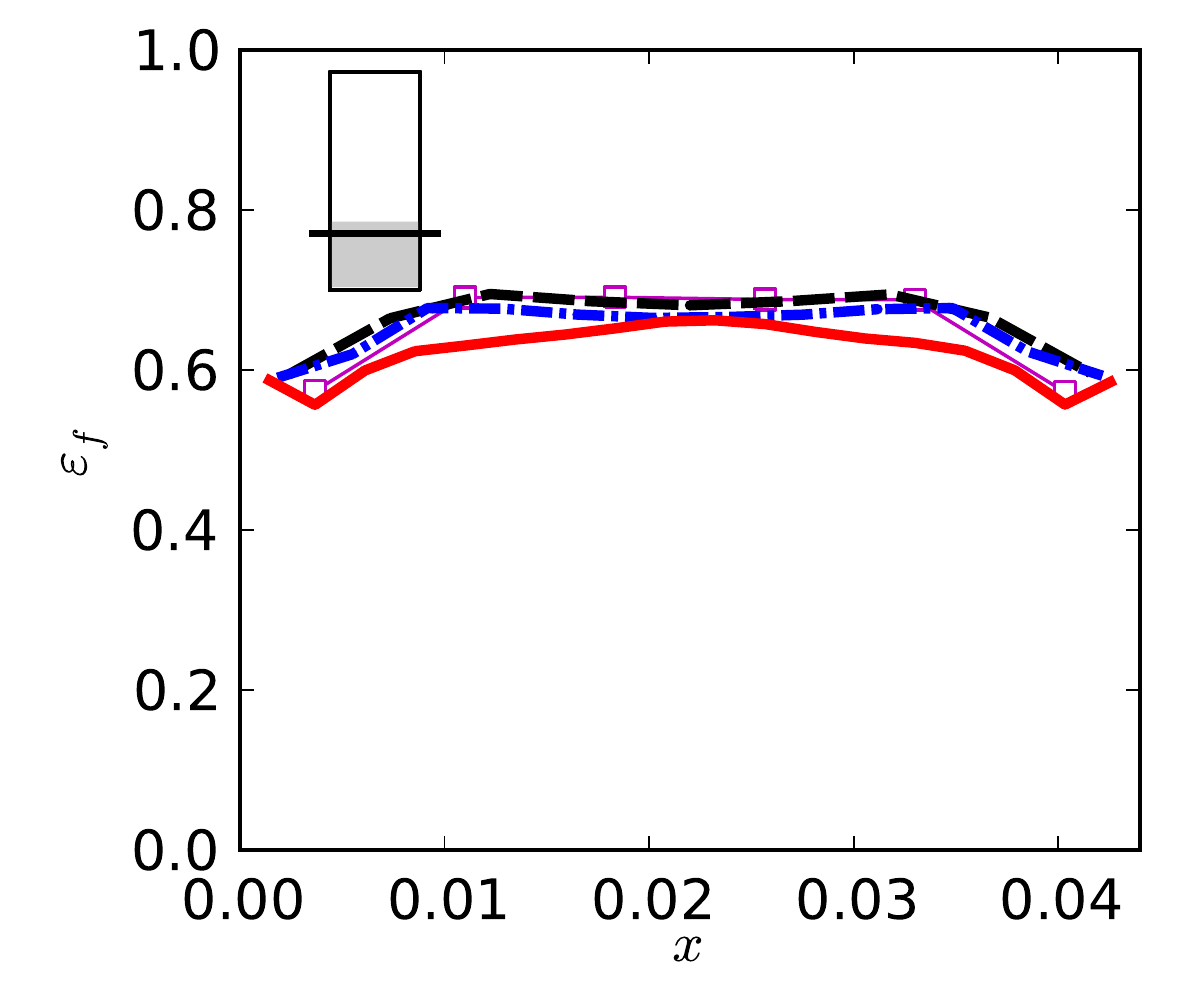}
  }
  \vspace{0.01\textwidth}
  \subfloat[][$x=11$~mm]{
    \includegraphics[width=0.45\textwidth]{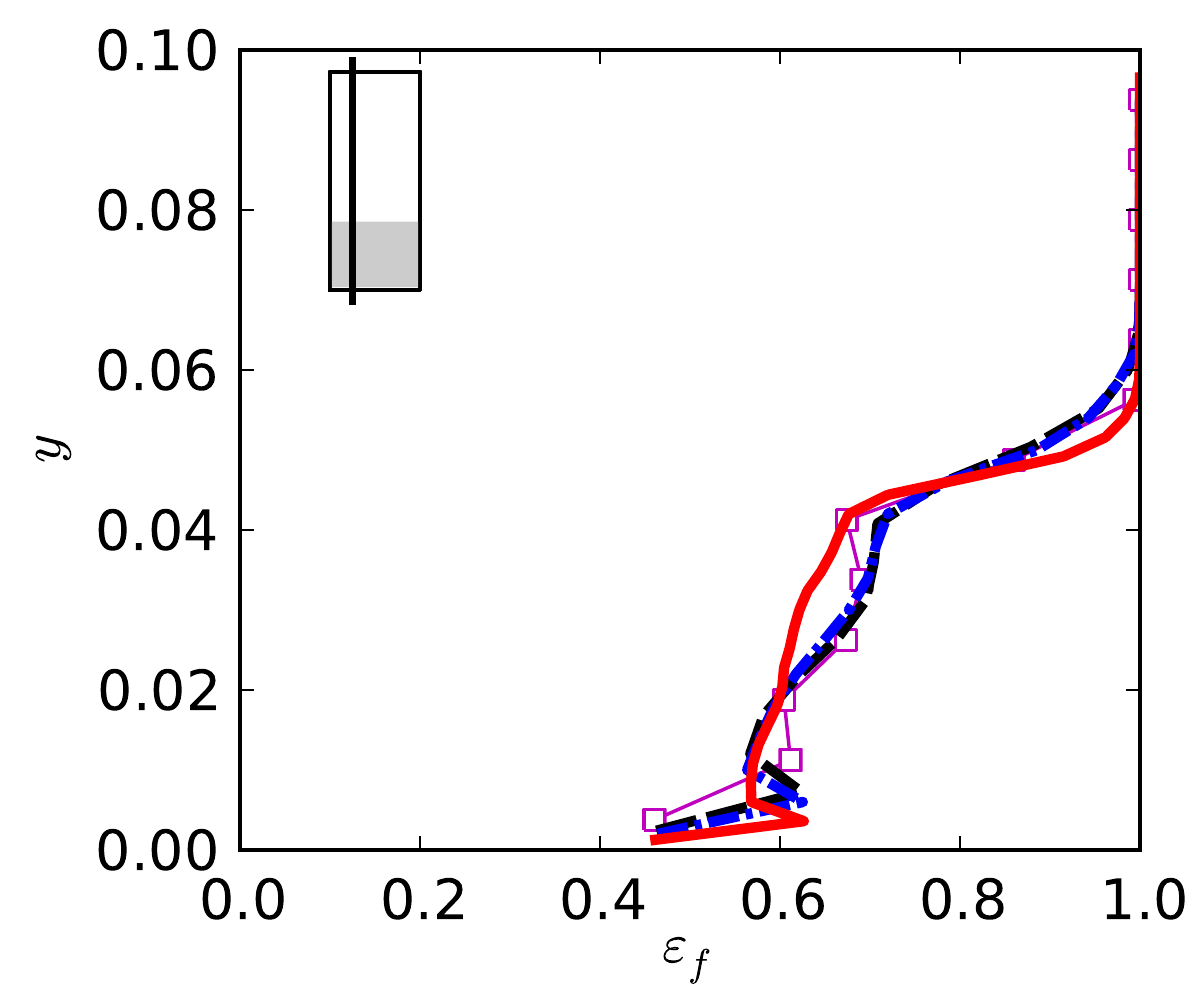}
  }
  \hspace{0.01\textwidth}
  \subfloat[][$x=22$~mm]{
    \includegraphics[width=0.45\textwidth]{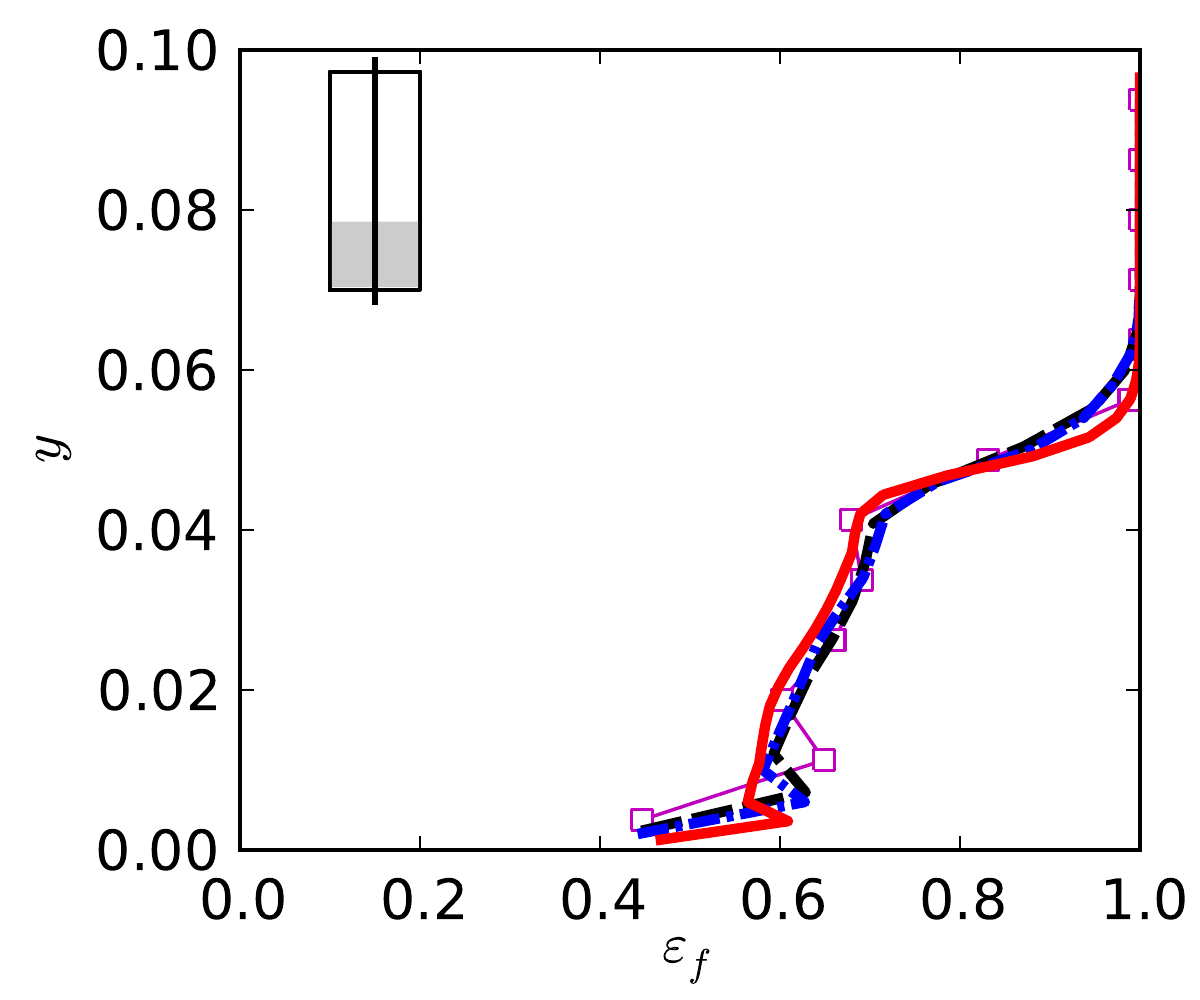}
  }
  \caption{Mesh convergence of PCM-based coarse graining, showing the profiles of fluid volume
    fraction $\varepsilon_f$ ($=1-\varepsilon_s$) on cross-sections at (a) $y=16.4$~mm, (b)
    $y=31.2$~mm, (c) $x=11$~mm, and (d) $x=22$~mm. Results from four meshes A, B, B$^\prime$, and C
    are compared. Refer to Fig.~\ref{fig:voidageIndependence} for detailed caption and for
    comparison.}
  \label{fig:voidageCompPCM}
\end{figure}

\begin{figure}[!htpb]
  \centering
  \subfloat[][$y=15$~mm]{
    \includegraphics[width=0.45\textwidth]{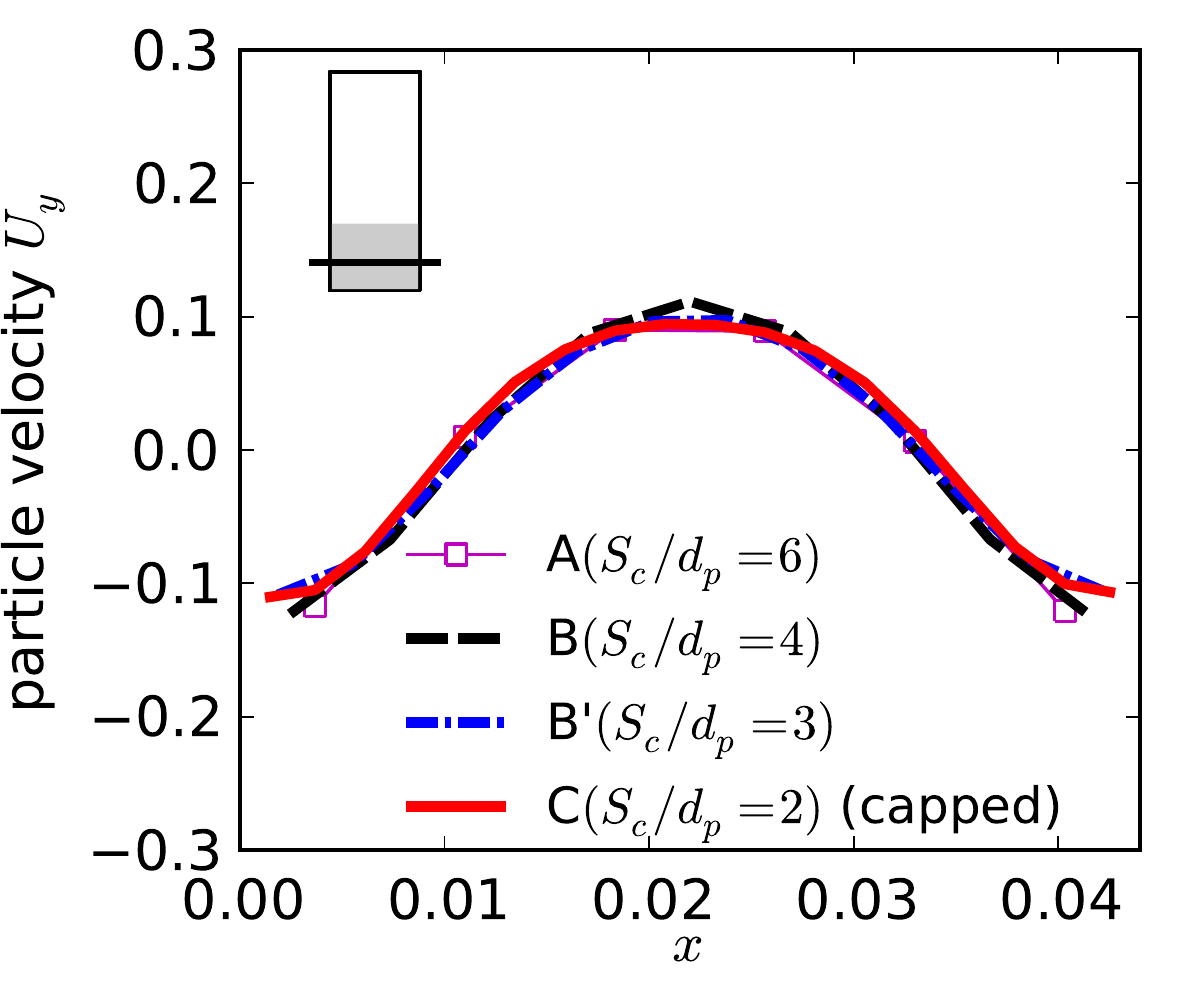}
  }
  \hspace{0.01\textwidth}
  \subfloat[][$y=25$~mm]{
    \includegraphics[width=0.45\textwidth]{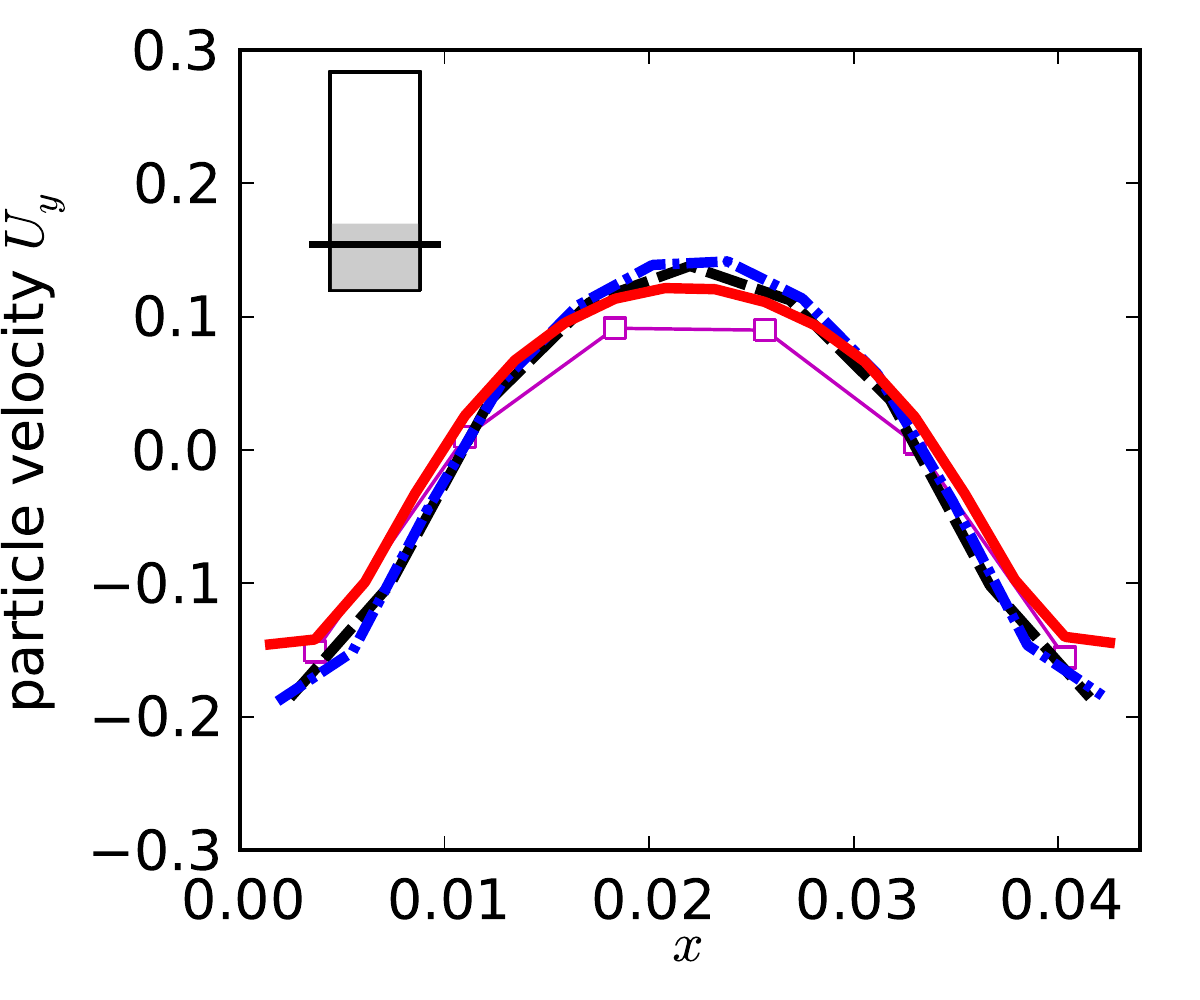}
  }
  \vspace{0.01\textwidth}
  \subfloat[][$x=11$~mm]{
    \includegraphics[width=0.45\textwidth]{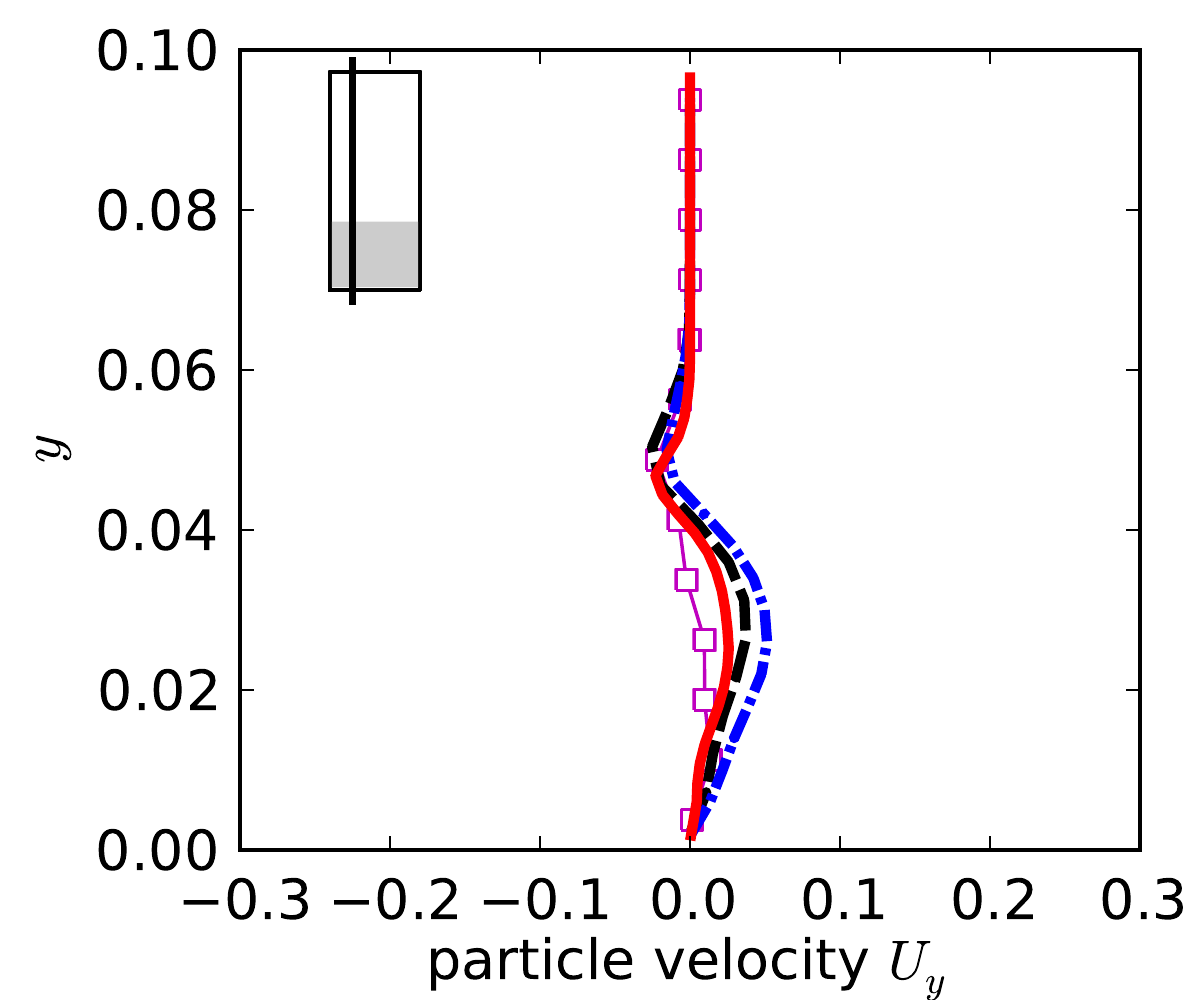}
  }
  \hspace{0.01\textwidth}
  \subfloat[][$x=22$~mm]{
    \includegraphics[width=0.45\textwidth]{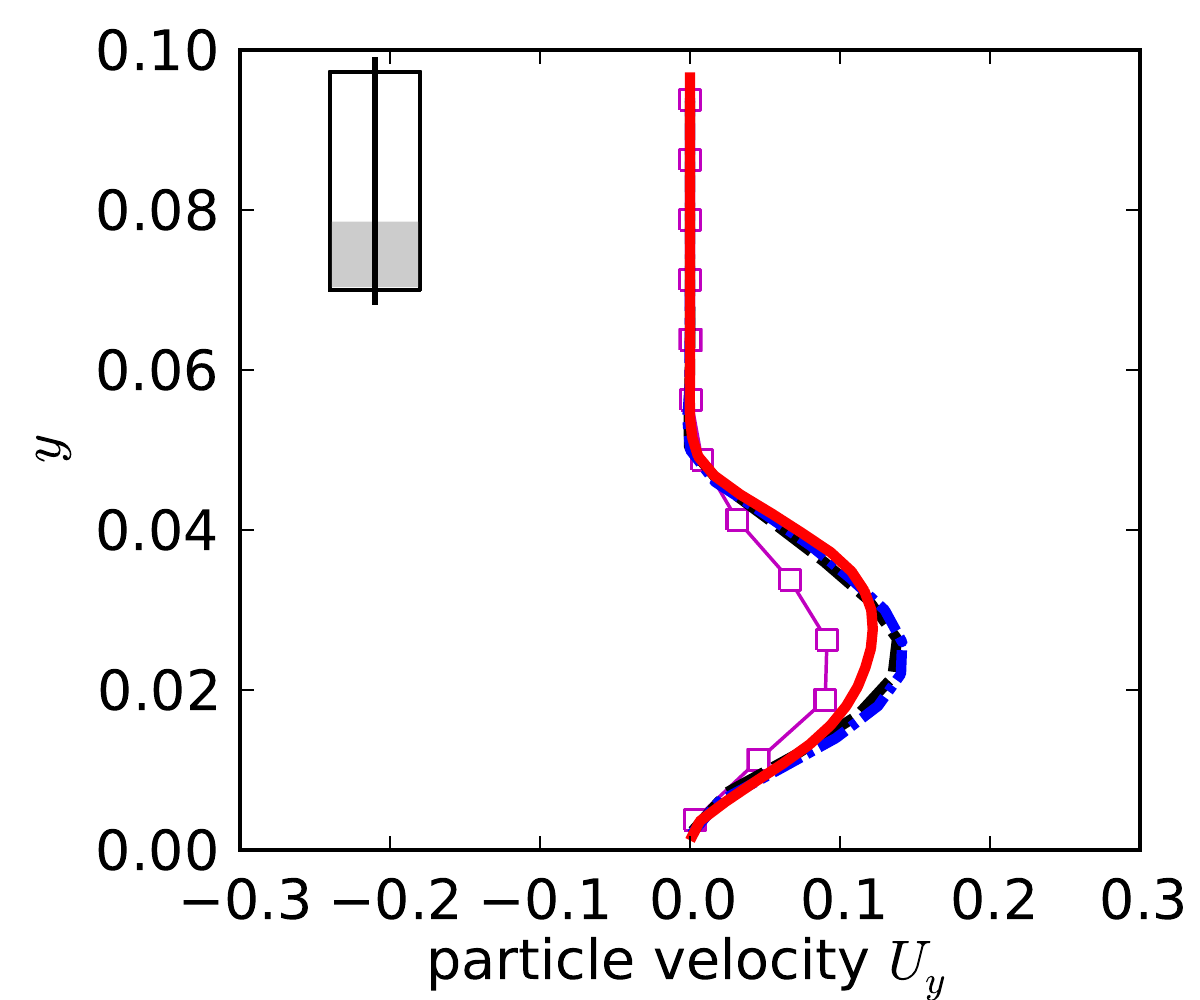}
  }
  \caption{Mesh convergence of PCM-based coarse graining, showing the profiles of vertical component
    $U_y$ of the time-averaged particle velocity at (a) $y=15$~mm, (b) $y=25$~mm, (c) $x=11$~mm, (d)
    $x=22$~mm.  Results from four meshes A, B, B$^\prime$, and C are compared. Refer to
    Fig.~\ref{fig:velocityIndependence} for detailed caption and for comparison.}
  \label{fig:velocityCompPCM}
\end{figure}

The profiles of time-averaged fluid volume fraction $\varepsilon_f$ obtained by using the same
CFD--DEM solver with PCM-based coarse-graining method are shown in
Fig.~\ref{fig:voidageCompPCM}. Profiles at the same four cross-section as in
Fig.~\ref{fig:voidageIndependence} are presented.  As can be seen from
Fig.~\ref{fig:voidageCompPCM}, at all cross-sections the $\varepsilon_f$ profiles obtained with
meshes B ($S_c/d_p = 4$) and B$^\prime$ ($S_c/d_p = 3$) are the same, indicating an approximate
mesh-independence. As with the results from the diffusion-based coarse-graining in
Fig.~\ref{fig:voidageIndependence}, the results obtained with mesh A ($S_c/d_p=6$) have some
discrepancies with the other results due to the inadequate mesh resolution. While this is not
evident in Figs.~\ref{fig:voidageCompPCM}(a) and \ref{fig:voidageCompPCM}(b), it can be seen at
several locations in Figs.~\ref{fig:voidageCompPCM}(c) and \ref{fig:voidageCompPCM}(d).  Note that
the deviations do not necessarily occur near the bottom boundary as in
Figs.~\ref{fig:voidageIndependence}(c) and (d), but at random locations instead. Perhaps the most
striking difference in the results in Fig.~\ref{fig:voidageCompPCM} is that the convergence is not
achieved when the mesh is further refined, as is evident from the fact the results of mesh C
($S_c/d_p = 2$) are different from those of meshes B and B$^\prime$.  With mesh C, unphysically large
$\varepsilon_s$ values are frequently encountered during the simulations, which cause
instabilities.  To address this issue associated with the PCM, a frequently used technique that is
adopted here is ``capping'', i.e., for all cells with solid volume fraction $\varepsilon_s$ larger
than a certain threshold value, e.g., $\varepsilon_{\textrm{threshold}} = 0.7$, the $\varepsilon_s$
in these cells are capped to be $\varepsilon_{\textrm{threshold}}$. This is done at each fluid time
step after the $\varepsilon_s$ field is calculated.  This technique improves the robustness of
CFD--DEM simulations with PCM-based coarse graining, but it may impair the accuracy of the fluid
drag calculation and the accuracy of the entire simulation, as an artificially set $\varepsilon_s$
value is used instead of the physical values in these cells. The capping technique may have caused
the failure of mesh-convergence observed in Fig.~\ref{fig:voidageCompPCM}.  The particle phase
velocities obtained on meshes A, B, B$^\prime$, and C with the PCM-based solver are presented in
Fig.~\ref{fig:velocityCompPCM}. Similar to what was observed in Fig.~\ref{fig:voidageCompPCM},
mesh-convergence is not achieved here, mainly due to the fact that mesh C produces result different
from those of meshes B and B$^\prime$. The discrepancies are most evident from
Figs.~\ref{fig:velocityCompPCM}(c) and (d).  In Fig.~\ref{fig:velocityCompPCM}(d), it can be seen
that even the results obtained with meshes B and B$^\prime$ are different.

\begin{figure}[!htpb]
  \centering
  \subfloat[][Diffusion-based method]{
    \includegraphics[width=0.9\textwidth]{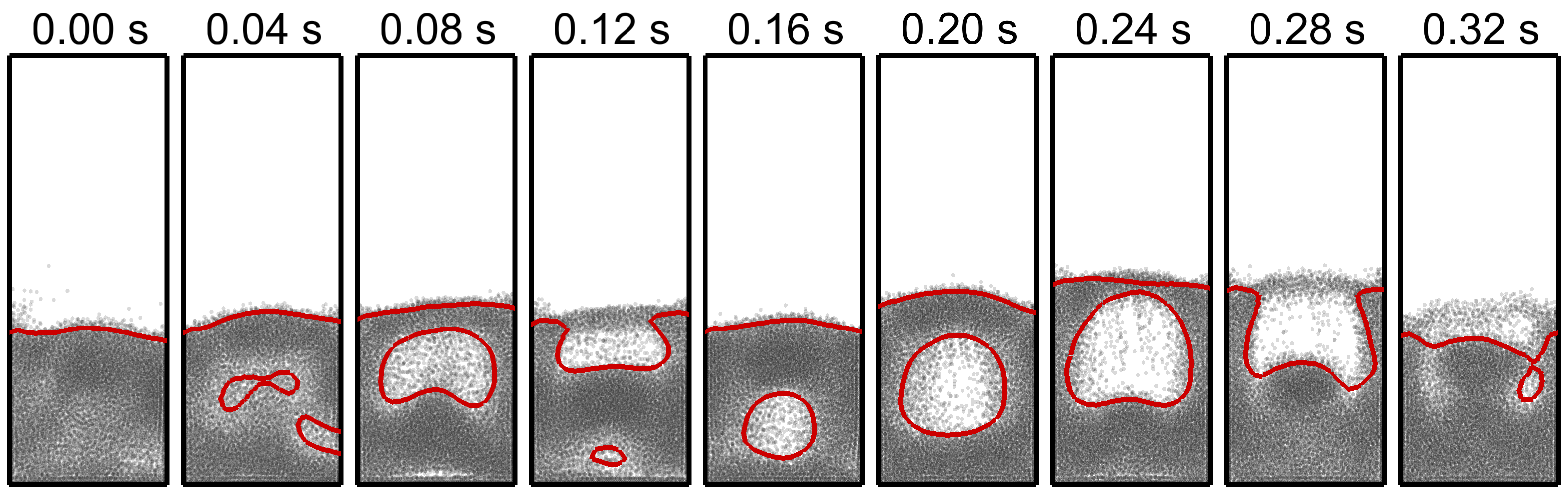}
}
  \vspace{0.02\textwidth}
  \subfloat[][PCM]{
    \includegraphics[width=0.9\textwidth]{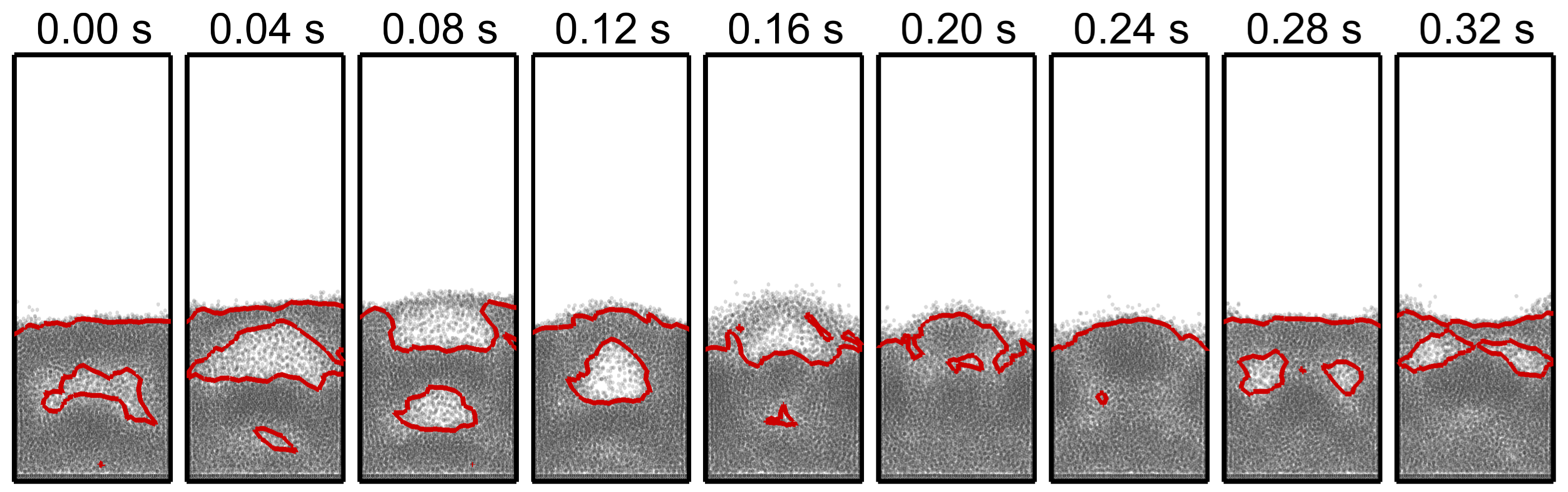}
  }
  \caption{
  \label{fig:voidageSnapshots}
  Snapshots of particle locations during a cycle of bubble formation and evolution, showing the
  results obtained by using the CFD--DEM solver with (a) diffusion-based and (b) PCM coarse-graining
  methods. The overlaying contours corresponds to $\varepsilon_s = 0.28$, which is half of the
  maximum $\varepsilon_s$ value in the initial bed configuration.  The time origin ($t=0$ s)
  corresponds to the beginning of the cycle.  }
\end{figure}

\begin{figure}[!htpb]
  \centering
  \subfloat[][Diffusion-based method]{
    \includegraphics[width=0.9\textwidth]{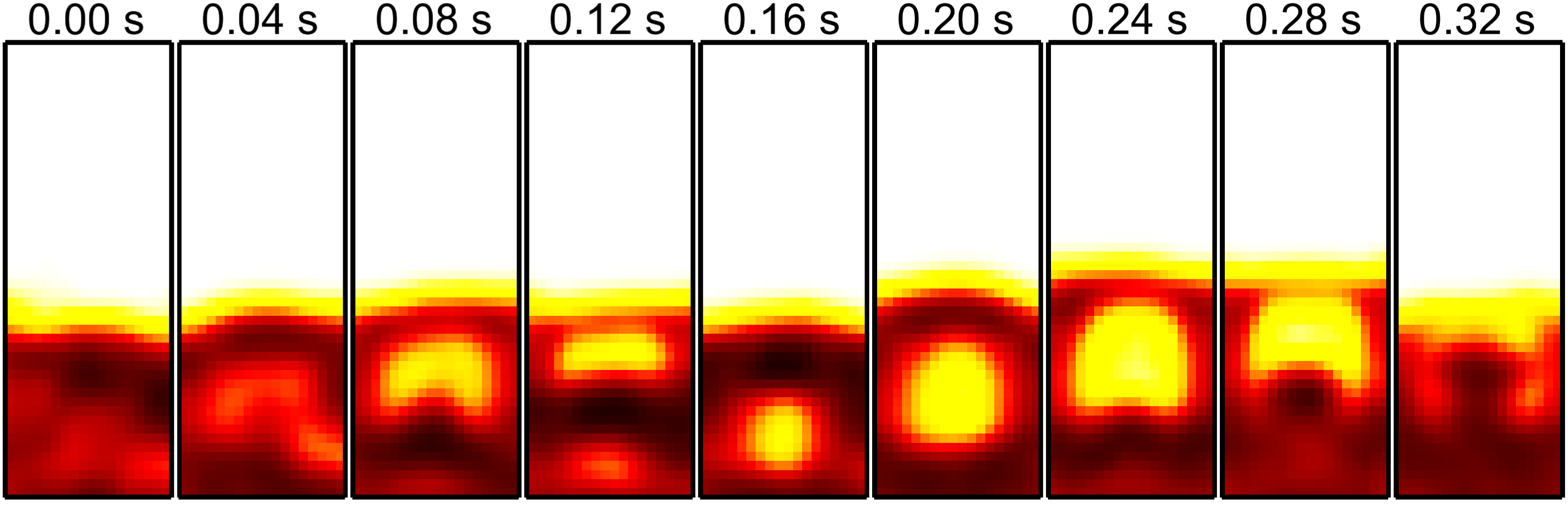}
}
 \vspace{0.02\textwidth}
  \subfloat[][PCM]{
 \includegraphics[width=0.9\textwidth]{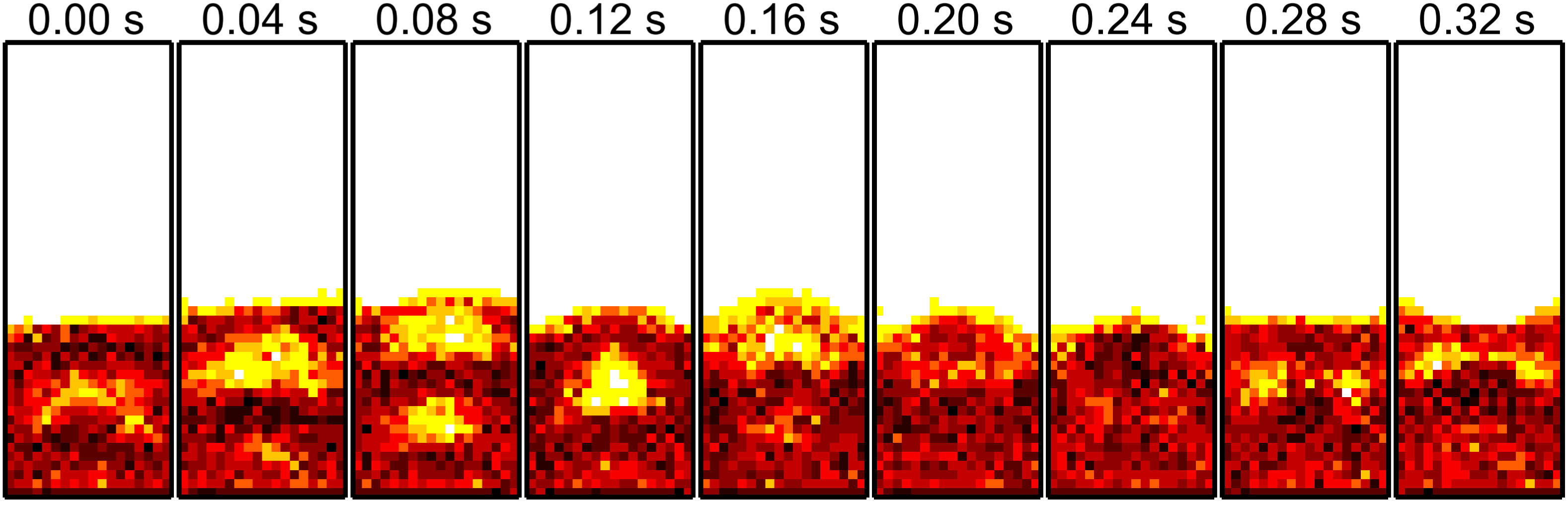}
  }
  \hspace{-0.0420\textwidth}
  \subfloat{
    \includegraphics[width=0.4\textwidth]{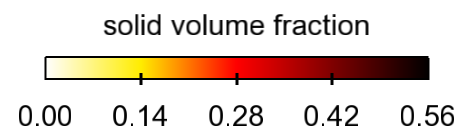}
  }
  \caption{
  \label{fig:fractionSnapshots}
  Snapshots of solid volume fraction $\varepsilon_s$ during a cycle of bubble formation and
  evolution, showing the results obtained by using the CFD--DEM solver with (a) diffusion-based and
  (b) PCM coarse-graining methods.  The time origin ($t=0$ s) corresponds to the beginning of the
  cycle.  }
\end{figure}

To further compare the different effects of the two coarse-graining methods on the CFD--DEM results,
snapshot sequences of particle locations during a bubble evolution cycle are presented in
Fig.~\ref{fig:voidageSnapshots}. The results from mesh C ($S_c/d_p=2$) are presented
here, since the mesh-convergence study above seems to suggest this to be a suitable mesh for both
cases.  The snapshots at $t=0.00$~s correspond to the beginning of the cycle. The volume fraction
contours corresponding to $\varepsilon_s = 0.28$ are overlaid on top of the particle location plots
to separate regions of lower and higher solid volume fractions, which facilitate visualization of
the bubble shapes and locations.  This contour value 0.28 is chosen to be one half of the maximum
solid volume fraction in the initial bed configuration.  From Fig.~\ref{fig:voidageSnapshots}(a),
the bubble formation ($t=0.04$ s), growth ($0.08$~s), and burst ($0.12$~s) can be clearly
identified.  At $t=0.12$~s, when the upper bubble bursts, a small bubble formed near the bottom of
the bed. In the subsequent snapshots from $0.16$~s to $0.24$~s, the bubble rises and grows until it
reaches the top of the bed at $0.28$~s, and bursts at $0.32$s eventually.  The bubble dynamics
observed here is physically reasonable as confirmed in previous experiments~\citep{muller08gr} and
numerical simulations~\citep{peng14in}.  In contrast, in the results obtained by using the same
CFD--DEM solver but with PCM-based coarse-graining, which is shown in
Fig.~\ref{fig:voidageSnapshots}(b), the cycle of bubble formation, growth, and bursting are not
observed as clearly, although the sequence from 0.04~s to 0.12~s does vaguely show a similar bubble
evolution dynamics as in Fig.~\ref{fig:voidageSnapshots}(a). Moreover, the bubble shapes in the PCM
results are much more irregular than those observed in the diffusion-based results.  To further
illustrate the different bubble dynamics, snapshots of solid volume fractions of the same cycle are
shown in Fig.~\ref{fig:fractionSnapshots}(a) and (b) for the CFD--DEM solvers with diffusion-based
and PCM-based coarse graining, respectively.  In Fig.~\ref{fig:fractionSnapshots}(a) the same
sequence of bubble dynamics as explained above are observed. The bubbles can be clearly identified
from the snapshots of solid volume fractions.  On the other hand, in the PCM results shown in
Fig.~\ref{fig:fractionSnapshots}(b), the bubbles are not as easily identified, which is partly due
to the large local variations (i.e., spatial gradients) in the $\varepsilon_s$ field.
 
In summary, with the same mesh, computational setup, parameters, and CFD--DEM solver but only with
different coarse-graining methods, the two carefully designed test cases clearly suggest the
superiority of the diffusion-based method compared with the PCM-based method in obtaining CFD--DEM
simulation results with mesh-convergence and in capturing the physics of the bubble dynamics in the
fluidized bed.

\subsection{Computational Overhead of Diffusion-Based Coarse Graining}
\label{sec:overhead}
In a typical CFD--DEM solver, the computational cost of DEM simulation dominates due to the high
computational costs in resolving particle collisions and the small time-scales in these
particle-scale processes.  In contrast, the number of cells in the CFD mesh and the associated
computational cost for the CFD simulation are generally moderate, since the cells must be large
enough to contain a sufficient number of particles for the locally averaged Navier--Stokes equations
to be valid.  Since the diffusion equations are solved on the CFD mesh, the computational cost
associated with the diffusion-based coarse graining is of the same order as that of the CFD
simulation.  Therefore, the additional computational costs incurred by solving the diffusion
equation are unlikely to be a significant portion of the entire CFD--DEM simulation. In LES--DEM,
however, the number of cells in the LES mesh may be large, and thus the computational overhead due
to the coarse-graining procedure based on solving diffusion equations may become a concern.  To
minimize the computational overhead, an implicit time stepping scheme should be used to guarantee
stability, which allows for large time step sizes to be used in the solution of the coarse-graining
diffusion equations. In~\cite{part1} we have shown that with any reasonable bandwidth the
diffusion equation can be solved with one time step to obtain sufficient accuracy, although some
minor fluctuations are still present in the coarse-grained field. If the diffusion equation is
solved with three time steps, the obtained field is sufficiently smooth, at least compared with
those obtained by using PCM and DPVM.

To investigate the computational costs associated with different parts of a typical CFD--DEM
simulation, a series of four numerical simulations are performed with the ratio~$N_p/N_c$ between
the number $N_p$ of particles and the number $N_c$ of CFD cells varying from 2 to 16.  In these
cases the number of particles~$N_p$ is kept constant at $3.32\times10^5$, while the numbers of CFD
cells~$N_c$ vary from $2.1 \times 10^4$ to $1.7 \times 10^5$.  The computational cases are
constructed according to those presented in Section~\ref{sec:insitu}, except that the computational
domain is enlarged to $264$~mm $\times~120$~mm $\times~60$~mm (in the width $x$-, height $y$-, and
transverse thickness $z$-directions, respectively) to allow for a realistically large number of
cells and particles to be used.  As in the experiment, the initial bed height is set to $30$~mm to
retain the same bed dynamics as in the experiment, but note that both the width and the transverse
thickness have been increased by six times. Since the CPU time needed for each time step is
relatively constant throughout the entire simulation, only 100 fluid time steps are simulated in
these tests.

The CPU times consumed by different parts of the CFD--DEM simulations, including the CFD part, the
DEM part, and the coarse graining part (mainly the solution of the diffusion equations), are
presented in Table~\ref{tab:insitu-cost} for the four cases studies. It can be seen from the table
that the total time spent on the DEM part do not vary much among the four cases, which is expected
since the total number of particles are the same in all the cases. On the other hand, the time spent
on the CFD part and that on the coarse graining increases as the number of CFD cells increases, at a
rate slightly higher than linear. It is worth noting that in all these cases the time spent on
coarse graining is slightly longer than the CFD part.  This is because the coarse graining diffusion
equations are performed for seven fields associated with three variables $\varepsilon_s$, $u_i$,
$F_i$ in each fluid time step. Three times steps are used each time when these equations are solved.
However, note that the percentage of CPU time spent on the coarse graining decreases with increasing
ratio $N_p/N_c$. Even for case 1 with $N_p/N_c = 2$, which indicates that there are only two
particles for each cell on average, the coarse graining accounts for only 28\% of the total
computational cost. Considering that the number of particles relative to the number of cells is very
small, this percentage is not really discouraging. \citet{peng14in} pointed out that a cell size of
$1.63d_p$ is needed to satisfy the CFD--DEM governing equations. A rule of thumb for CFD--DEM
simulations is that on average a typical cell should contain approximately nine
particles~\citep{muller09va,xiao-cicp}. In view of these observations, cases 3 and 4 are probably
more realistic in terms of $N_p/N_c$ ratios. In the two cases, the diffusion-based coarse-graining
procedure accounts for only 10\% and 5\%, respectively, of the total cost of the CFD--DEM
simulations, which we believe are rather moderate. {\color{black} Admittedly, the diffusion equations
  incur more computational costs than the CFD solver. This is in stark contrast to PCM and DPVM,
  which are expected to incur negligible computational costs. The computational overhead of the
  diffusion-based method may be significant and undesirable when fine meshes are used, i.e., when
  high-resolution solvers such as LES are used in the CFD part. In these cases, one may consider
  further increasing the time step size used in solving the diffusion equations, e.g., solving the
  diffusion equation in only one time step with an implicit scheme
  \citep[see][]{part1}. Alternatively, one can recast the continuity equation (\ref{eq:NS-cont}) to
  eliminate the term $\varepsilon_s \mathbf{U}_s$ (see Eq. (1a) in \citet{wu14parallel}), so that
  the number of diffusion equations to solve is reduced from seven to four.}

A few additional observations need to be made when examining the computational overhead due to the
diffusion-based coarse-graining procedure. First, we have demonstrated in the companion
paper~\citep{part1} that even on unfavorable meshes the diffusion-based coarse-graining method lead
to smooth solid volume fraction $\varepsilon_s$ fields (as well as $\mathbf{U}_s$ and
$\mathbf{F}^{fp}$). As observed by~\cite{peng14in}, compared to the non-smooth coarse grained fields
(e.g., $\varepsilon_s$) such as those obtained by using the PCM, a smooth $\varepsilon_s$ field can
significantly accelerate convergence in the CFD solver, and thus effectively reduces overall
computational costs of the CFD--DEM simulation. Second, the parallelization of the diffusion-based
method is straightforward and can very effectively take advantage of exiting infrastructure in the
CFD solver, which is often highly optimized. The two factors partly offset the moderate
computational overhead associated with the diffusion-based coarse-graining procedure. {\color{black}
  On the other hand, as demonstrated in previous studies~\citep{wu06dense,wu09three, xiao-cicp},
  linearization and implicit treatment of the fluid--particle momentum exchange terms are effective
  methods for accelerating the convergence of the PISO algorithm by increasing the diagonal
  dominance in the matrix of the discretized linear equations systems. Diffusing the fluid--particle
  drag forces as performed in the proposed method makes it difficult, if not impossible, to perform
  such linearizations. This is a potential limitation of the diffusion-based coarse-graining method
  if linearization and implicit treatment are essential, e.g., in such challenging cases as granular
  Rayleigh-Taylor instability problems~\citep{wu14parallel}.}

\begin{table}[!htbp]
  \caption{
  \label{tab:insitu-cost}
  Breakdown of computational costs associated with different parts of CFD--DEM
  simulations. The computational costs are presented for four cases with the same number of  
  particles $N_p=3.3\times 10^{5}$ and different numbers $N_c$
  of CFD cells. The CPU times presented here are normalized by the time spent on the CFD part of case~4,
  which has the smallest number of CFD cells.
}
  \begin{center}
  \begin{tabular}{cccccc}
    \hline
    case & $N_c$ & $N_p/N_c$ & CFD  & DEM & coarse graining\\
    \hline
    1 & $1.7 \times 10^5$ & 2  & $12~(23\%)$  & $26~(49\%)$ & $14.8~(28\%)$  \\
    2 & $8.3 \times 10^4$ & 4  & $~5~(13\%)$  & $27~(70\%)$ & $~6.4~(17\%)$  \\
    3 & $4.1 \times 10^4$ & 8  & $~2~(~7\%)$  & $25~(83\%)$ & $~2.8~(10\%)$  \\
    4 & $2.1 \times 10^4$ & 16 & $~1~(~3\%)$  & $27~(92\%)$ & $~1.5~(~5\%)$  \\
    \hline
  \end{tabular}
  \end{center}

\end{table}

\section{Discussion}
\label{sec:discuss}

\subsection{Choice of Bandwidth in Kernel Functions for Coarse Graining}
\label{sec:choose-b}

\begin{figure}[!htbp]
\centering	
  \includegraphics[width=0.6\textwidth]{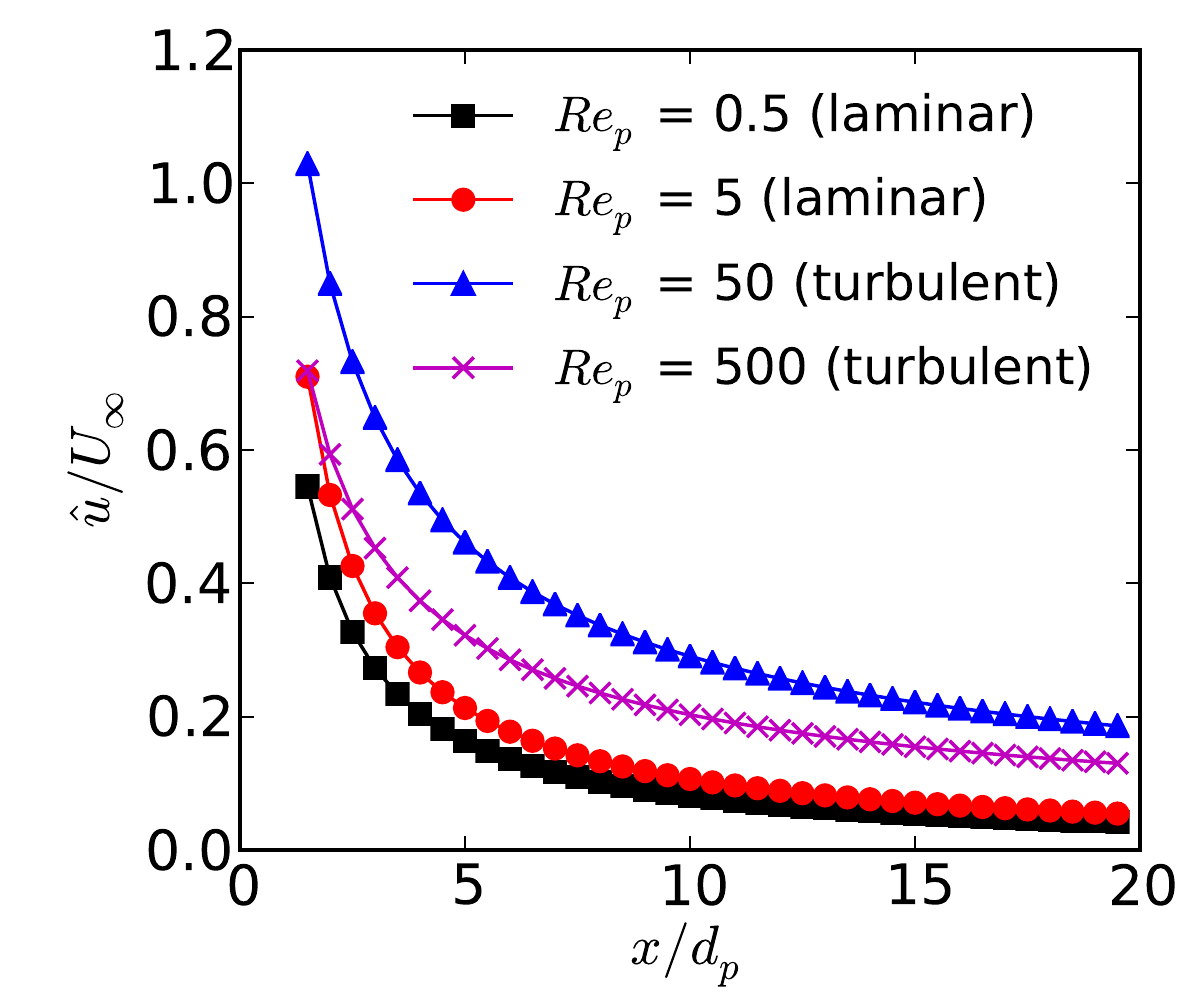}
  \caption{Velocity defect $\hat{u} \equiv U_\infty - u$ normalized by the free-stream velocity
    $U_\infty$ in the wake of a spherical particle along the imaginary streamline passing through
    the center of the sphere, illustrating the size of the wake of a particle at different flow
    regimes and particle Reynolds numbers, which are defined as $Re_p = U_\infty d_p/\nu_f$ (where
    $\nu_f$ is the fluid kinematic viscosity). Results are obtained according to the empirical
    formulas in~\cite{wu93sphere} and are presented for a range of typical particle Reynolds
    numbers.  To obtain these $Re_p$ values, particle--fluid relative velocities $U_\infty$ ranging
    from 0.005~m/s to 5~m/s, a particle diameter of $d_p = 1$~mm, and a fluid viscosity of $\nu_f =
    1.0\times10^{-6}~\mathrm{m^2/s}$ are used.}
\label{fig:wu-faeth}
\end{figure}

In coarse-graining procedures used to link microscopic and macroscopic quantities, the choice of
parameters (e.g., the bandwidth $b$ in the kernel functions) remains an open
question~\citep{latzel00macro,zhu02ave}. Here we argue that in CFD--DEM simulations the bandwidth
$b$ should be chosen based on the size of the wake of the particles in the fluid flow, which in
turn depends on the particle diameter and the particle Reynolds number, among other
parameters~\citep{wu93sphere}. While coarse graining may be performed for different theoretical and
practical purposes depending on the physical context,  in CFD--DEM  the main reason for the coarse
graining is to compute the interactions forces between the fluid and the particle phases.
Specifically, for example, the solid volume fraction \( \varepsilon_s \) is needed in the drag
force calculation to account for the effects of a particle on the drag experienced by other
particles. It is also needed in the continuity and momentum equations of the fluid phase. In the
latter,  in addition to \( \varepsilon_s \) and $\xmb{U}_s$, which are obtained via coarse
graining,  the force of a particle on the fluid needs to be distributed to an appropriate volume of
the fluid. In light of this observation, from a physical perspective the support of the kernel
function should extend approximately to the same distance as the wake of the particles. The wake
size depends on the particle Reynolds number, which in turn depends on the relative velocity between the
particle and the fluid. To compute the wake size of the particles (indicated by the velocity defect
along the particle centerline) in the case studied in Section~\ref{sec:insitu}, a few
representative relative velocity values are used, and the results are displayed in
Fig.~\ref{fig:wu-faeth}.

A bandwidth between $b=4d_p$ and $6d_p$ was used in the simulations in the current work
(Section~\ref{sec:insitu}) and in the \emph{a priori} tests in the companion
paper~\citep{part1}. Since the support of the kernel is approximately $3b$, this choice of bandwidth
corresponds to a kernel support of approximately 12 to 18 particle diameters.  This is indeed of the
same order of magnitude as that suggested by the wake sizes shown in Fig.~\ref{fig:wu-faeth}. In a
polydispersed system with a range of particle diameters, although changing bandwidth $b$ (or
equivalently diffusion time span $T$) adaptively according to relative particle velocities would be
unrealistic and probably unnecessary as well, it is possible and easy to accommodate different
particles sizes (and thus wake sizes) by using a spatially varying diffusion coefficient in
Eq.~(\ref{eq:diffusion}). Currently, a diffusion coefficient that is uniformly one is used
throughout the domain.

\subsection{Implementation of Time--Volume Averaging in CFD--DEM Solvers}
\label{sec:tv-avg}
As with most previous CFD--DEM works, in this study we consider only volume averaging in the coarse
graining procedure. That is, the kernel \( G(\xmb{x}) \) is a spatial function, and it operates on
particle distributions at a particular time only. Accordingly, the diffusion equation as in
Eq.~(\ref{eq:diffusion}) is solved only to smoothen the coarse-grained fields at each time fluid
step. Note that the time variable $\tau$ in the diffusion equation is the pseudo-time, and not the
physical time. In principle it is possible to include time in the kernel function as well, i.e., \(
G = G(\xmb{x}, t) \), with the normalization condition revised to be \( \int_{\mathbb{R}^4}
G(\xmb{x}, t) d\xmb{x} dt = 0 \). This is indeed what was proposed in~\cite{zhu02ave}. However,
the kernel function \( G(\xmb{x}, t) \) is symmetric both spatially and temporally. Essentially, the
solid volume fraction \( \varepsilon_s \) at any location is contributed to by particles in its
surroundings in all directions, with closer particles having more contributions. Similarly, \(
\varepsilon_s\) depends on the particle distributions at both past and future times. This spatial
and temporal symmetry is desirable from a theoretical point of view. However, in CFD--DEM solvers a
kernel function with the symmetry in time cannot be implemented, since at time $t$ when we need to
compute \( \varepsilon_s (x, t) \) the particle distributions at future times are not known but
need to be solved. If coarse graining or averaging in time is indeed desirable or necessary (e.g.,
when the flow fields have high-frequency fluctuations such as in LES--DEM), an alternative to obtain
a field with time--volume averaging, e.g., \( \left< \varepsilon_s (\xmb{x}, t) \right> \), is to
use a single-sided averaging scheme as follows~\citep{meneveau96,xiao12jcp}:
\begin{align}
\left< \varepsilon_s (\xmb{x}, t) \right> & = \int_0^t \varepsilon_s (\xmb{x}, t') W(t-t') dt'  \label{eq:cg-time} \\
	 \textrm{with} \quad W(t-t')  & = \frac{1}{b_t} \exp\left[-(t-t')/b_t \right], \notag
\end{align}
where \( \varepsilon_s \) is the solid volume fraction with volume averaging only, $W$ is the
exponential kernel function, and $b_t$ is the temporal averaging bandwidth with similar
interpretation to $b$ in Eq.~(\ref{eq:hi}). Compared with the Gaussian kernel function, a
convenient feature of the exponential kernel function is that \( \left< \varepsilon_s \right> \) as
defined in Eq.~(\ref{eq:cg-time}) is the solution of the following differential equation:
\begin{equation}
\label{eq:cg-time-eqn}
\frac{d \left<\varepsilon_s \right>}{dt} = \frac{1}{b_t} (\varepsilon_s  - \left<\varepsilon_s
\right>) ,
\end{equation}
which can be approximated to the first order by the following relation:
\begin{align}
\left<\varepsilon_s \right>^n & = (1-\alpha) \varepsilon_s^n + \alpha \left<\varepsilon_s \right>^{n-1}
\label{eq:cg-time-disretize}\\
\textrm{with} \quad \alpha  &= \frac{1}{1+\Delta t / b_t} ,
\end{align}
where $n$ and $n-1$ are the time indices of the present and the previous time steps, respectively.
The scheme in Eq.~(\ref{eq:cg-time-disretize}) suggests that to compute \( \left< \varepsilon_s
\right> \) for the current step, one only needs \( \varepsilon_s \) of the current step and \(
\left< \varepsilon_s \right> \) of the previous time step. This would lead to reduced storage
requirements compared with a literal implementation of Eq.~(\ref{eq:cg-time}), where \( \varepsilon_s
\) at many previous time steps need to be stored. While the above-mentioned characteristics of
exponential kernel function are exploited in the averaging in turbulence flow
simulations~\citep{meneveau96,xiao12jcp}, we are not aware of any such attempts in the context of
coarse graining in CFD--DEM simulations. Hence, we chose to present it here for the completeness of
the diffusion-based coarse-graining algorithm and for the reader's reference.

\section{Conclusion}
\label{sec:conclude}
In this work we applied the previously proposed coarse-graining algorithm based on solving diffusion
equations~\citep{part1} to CFD--DEM simulations.  The conservation requirements are examined and
satisfied by properly choosing variables to solve diffusion equations for.  Subsequently, the
algorithm is implemented into a CFD--DEM solver based on OpenFOAM and LAMMPS, the former being a
general-purpose, three-dimensional, parallel CFD solver based on unstructured meshes. The
implementation is straightforward, fully utilizing the computational infrastructure provided by the
CFD solver, including the parallel computing capabilities. Simulations of a fluidized bed showed
that the diffusion-based coarse-graining method led to more robust simulations, physically more
realistic results, and improved ability to handle small CFD cell-size to particle-diameter
ratios. Moreover, the mesh convergence characteristics of the diffusion-based method are
dramatically improved compared with the PCM. It is demonstrated in the current work that the
diffusion-based method lead to mesh-independent results in CFD--DEM simulations, confirming the
conclusions drawn from the \emph{a priori} tests in the companion paper~\citep{part1}. The choice of
the computational parameter, i.e., the bandwidth, has clear physical justifications.  {\color{black}
  The computational costs of the proposed method were carefully investigated.  Results suggest that,
  although the computational overhead due to the diffusion-based coarse graining exceeded that of
  the CFD solver in all cases, the additional computational costs are not significant (less than
  10\% of the total costs) if the number of particles per cell is large, i.e., when the computational
  costs of the simulations are dominated by the DEM part.}  Therefore, the diffusion-based method is
a theoretically elegant and practically viable option for coarse graining in general-purpose
CFD--DEM solvers.

\section{Acknowledgment}

The computational resources used for this project were provided by the Advanced Research Computing (ARC) of Virginia Tech, which is gratefully acknowledged. We thank the anonymous reviewers for  their comments, which helped improving the quality of the manuscript.

\bibliographystyle{elsarticle-harv}
% \bibliographystyle{model1-num-names}
% \bibliography{diffusionEqn,ACS-PRF,ACS-New}

\appendix
\section{Conservation Characteristics of the Diffusion-Based Method}
The conservation requirements for the particle mass, particle momentum, total momentum in the
fluid--particle system as specified in Section~\ref{sec:conserve} are summarized as follows:
 \begin{subequations}
  \label{eq:conserve}
  \begin{align}
   \rho_s\sum_{k = 1}^{N_c} \varepsilon_{s, k} \, V_{c, k} &
   = \sum_{i = 1}^{N_p} \rho_s \, V_{p, i} \; , \label{eq:conserve-eps} \\
   \rho_s \sum_{k = 1}^{N_c} \varepsilon_{s, k} \, V_{c, k} \, \mathbf{U}_{s, k} &
   = \sum_{i = 1}^{N_p} \rho_s \,V_{p, i} \, \mathbf{u}_{p, i} \; , \label{eq:conserve-u} \\
   \sum_{k = 1}^{N_c} (1-\varepsilon_{s, k}) \rho_f \, V_{c, k} \, \mathbf{F}^{fp}_k &
   = -\sum_{i = 1}^{N_p} \, \mathbf{f}^{fp}_i \; ,\label{eq:conserve-f}
  \end{align}
 \end{subequations}
 where the density \( \rho_s \) is assumed to be constant for all particles;
 \(N_c\) is the number of cells in the CFD mesh; \(N_p\) is the number of particles in the system;
 \(V_{c, k}\) is the volume of cell \(k\); \(\mathbf{U}_{s, k}\) is the Eulerian solid phase
 velocity in cell \(k\); \(\mathbf{u}_{p,i}\) is the Lagrangian velocity of particle \(i\);
 \(\mathbf{F}^{fp}_k\) is the force per unit fluid mass exerted on fluid cell \(k\) by all
 particles; \(\mathbf{f}^{fp}_i\) is the fluid force on particle \(i\).

 Multiplying both sides of the equations in the PCM coarse-graining procedure, i.e.,
 Eqs.~(\ref{eq:pcm-k}), ~(\ref{eq:pcm-u}), and (\ref{eq:pcm-f}), by \(V_{c, k}\), \(\rho_s
 \varepsilon_{s,k} V_{c, k}\), and \(\rho_f \varepsilon_{f,k} V_{c, k}\), respectively, and taking
 summation over all cells, the conservation requirements in Eq.~(\ref{eq:conserve}) can be
 recovered. Therefore, the PCM-based coarse-graining schemes as in Eqs.~(\ref{eq:pcm-k}),
 ~(\ref{eq:pcm-u}), and (\ref{eq:pcm-f}) are conservative \emph{by construction}. The proposed
 coarse-graining algorithm consists of two steps: (1) coarse graining using PCM; and (2) solving
 diffusion equations for the quantities $\varepsilon_s$, $\varepsilon_s \mathbf{U}_s $ and
 $\varepsilon_f \mathbf{F}^{fp}$. Hence, for the proposed algorithm to be conservative, the
 diffusion step must also conserve the required physical quantities in the domain, i.e.,
\begin{subequations}
 \label{eq:conserve-qs}
 \begin{align}
  \textrm{particle mass}: \quad &
  \rho_s\sum_{k = 1}^{N_c} \varepsilon_{s, k} \, V_{c, k} \; , \\
  \textrm{particle momentum}: \quad &
  \rho_s \sum_{k = 1}^{N_c} \varepsilon_{k} \, V_{c, k} \, \mathbf{U}_{s, k} \; , \; \\
  \textrm{total momentum of fluid--particle system}: \quad &
\rho_f  \sum_{k = 1}^{N_c} \varepsilon_{f, k} \,  V_{c, k} \, \mathbf{F}^{fp}_{k} \; .
 \end{align}
\end{subequations}

For the diffusion equation \({\partial \phi}/{\partial t} = \nabla^2 \phi\) of a generic variable
\( \phi \) on domain \(\Omega\) with no-flux condition \(\partial \phi / \partial n = 0\) on the
boundaries \(\partial \Omega\), integrating the equation on $\Omega$ yields:
\begin{subequations}
 \begin{align}
  \int\limits_{\Omega} \frac{\partial \phi}{\partial t} d\Omega
  = \int\limits_{\Omega} \nabla^2 \phi d\Omega = \int\limits_{\partial \Omega} \nabla \phi \cdot d\mathbf{S} & = 0,  \\
\textrm{or equivalently,} \quad  \frac{\partial}{\partial t} \int\limits_{\Omega} \phi d\Omega &
  = 0,
 \end{align}
\end{subequations}
which suggests that the conserved quantity that is implied by the diffusion equation
\({\partial \phi}/{\partial t} = \nabla^2 \phi\) is
\begin{equation}
  \label{eq:conserved-phi}
 \int_{\Omega} \phi \, d\Omega ,
\end{equation}
 or
\begin{equation}
 \sum_{k=1}^{N_c} \phi_i V_{c, k}
 \label{eq:phi-vc}
\end{equation}
on a discretized finite volume mesh. Comparing Eq.~(\ref{eq:phi-vc}) with the quantities in
Eq.~(\ref{eq:conserve-qs}) suggests that diffusion equations should be solved for the following
three quantities:
\begin{equation}
\varepsilon_s, \quad
 \varepsilon_s \mathbf{U}_s , \quad
\textrm{ and } \quad
 \varepsilon_f  \mathbf{F}^{fp}
\end{equation}
to satisfy the conservation requirements for particle mass, particle momentum, total momentum in the
fluid--particle system, respectively. 

\end{document}